\documentclass[%
reprint,
 amsmath,amssymb,
 aps,
pre,
floatfix,
]{revtex4-2}
\usepackage{graphics}
\usepackage{dcolumn}
\usepackage{bm}
\usepackage{graphicx}
\usepackage{amsmath}
\usepackage{amssymb}
\usepackage{float}
\usepackage[caption=false]{subfig}
\usepackage{xcolor}
\usepackage{mathtools}
\usepackage{booktabs}

\newcommand{\be}{\begin{equation}}
\newcommand{\ee}{\end{equation}}
\newcommand{\bea}{\begin{eqnarray}}
\newcommand{\eea}{\end{eqnarray}}

\begin{document}

\title{Parameter study of decaying magnetohydrodynamic turbulence }
\author{Andres Armua}
\email{andres.armua@ed.ac.uk}
\author{Arjun Berera}
\email{ab@ph.ed.ac.uk}
\author{Jaime Calder\'on-Figueroa}
\email{jaime.calderon@ed.ac.uk}
\affiliation{
School of Physics and Astronomy,
University of Edinburgh, Edinburgh EH9 3FD, United Kingdom}

\begin{abstract}
It is well known that helical magnetohydrodynamic (MHD) turbulence exhibits an inverse transfer of magnetic energy from small to large scales, which is related to the approximate conservation of magnetic helicity. Recently, several numerical investigations noticed the existence of an inverse energy transfer also in nonhelical MHD flows. We run a set of fully resolved direct numerical simulations and perform a wide parameter study of the inverse energy transfer and the decaying laws of helical and nonhelical MHD. Our numerical results show only a small inverse transfer of energy that grows as with increasing Prandtl number (Pm). This latter feature may have interesting consequences for cosmic magnetic field evolution. Additionally, we find that the decaying laws $E \sim t^{-p}$ are independent of the scale separation and depend solely on Pm and Re. In the helical case we measure a dependence of the form $p_b \approx 0.6 + 14/$Re. We also make a comparison between our results and previous literature and discuss the possible reason for the observed disagreements. 
\end{abstract}

\date{\today}

\maketitle

\section{Introduction}
\label{se:intro}

Decaying magnetohydrodynamic (MHD) turbulence received special attention in recent years. The study of decaying turbulence has been a topic of interest in its own sake for decades  \cite{saffman1967large,batchelor1948decay,kolmogorov1941degeneration,lee2010lack,christensson2001inverse,herring1974decay,biskamp1999decay,davidson1997role,davidson2000loitsyansky,davidson2010decay}. Furthermore, the decay of MHD turbulence is of central importance for astrophysics and cosmology, especially for the generation and evolution of large scale cosmic magnetic fields \cite{banerjee2004evolution,subramanian2016origin,campanelli2014evolution,mtchedlidze2021evolution,hosking2022cosmic}. 

The presence of an inverse cascade in helical MHD turbulence has been studied for many decades now \cite{frisch1975possibility}, and a large number of numerical studies have been dedicated to this topic. Some used direct numerical simulations (DNS) \cite{alexakis2006inverse,stepanov2014joint,brandenburg2001inverse,meneguzzi1981helical,balsara1999formation,christensson2001inverse,muller2012inverse}, whereas others used closure approximations and cascade models \cite{brandenburg1996large,brandenburg1997effect,pouquet1976strong}. 

In recent years, several studies found evidence of an inverse energy transfer also in nonhelical flows \cite{mckay2019fully,brandenburg2015nonhelical,berera2014magnetic,reppin2017nonhelical,bhat2021inverse,hosking2021reconnection,park2017inverse,zrake2014inverse}. The physical mechanisms involved in the nonhelical inverse transfer are different to those in the case of magnetic helicity, and are not completely understood yet. Some recent studies claim that magnetic reconnection may play an essential role in this inverse transfer \cite{park2017inverse,zhou2019magnetic,zhou2020multi,zhou2021statistical,bhat2021inverse,hosking2021reconnection, zhou2022scaling}. Only a few works analyzed the dependence of this effect on the magnetic Prandtl number Pm $= \nu/\eta$, which is the ratio of kinetic viscosity to magnetic diffusivity \cite{reppin2017nonhelical,mckay2019fully}. This is important for applications, since it is estimated that Pm $\gg 1$ in astrophysical systems such as the interstellar and intergalactic medium   \cite{schober2012magnetic,verma2004statistical,kulsrud1999critical}. This is also noted in \cite{park2017inverse}, where the authors choose a value of Pm $\gg1$.

In this work we perform a series of fully resolved DNS of decaying MHD turbulence with special focus on the inverse transfer of magnetic energy as well as the decaying exponents for a wide range of parameters such as Prandtl number, Reynolds number and scale separation. Most of the simulations in this paper are initially in equipartition and a small number are magnetically dominated. This is contrary to most recent numerical results which are mainly magnetically dominated.

Recent numerical studies implement hyperviscosity and hyperresistivity to overcome resolution limitations. Our whole analysis is based on flows with standard viscosity and we only run a few hyperviscous runs for the sake of comparison.   

The paper is organized as follows. In section \ref{se:decaying_turbulence} we give a brief introduction to decaying hydrodynamics, decaying MHD for both helical and nonhelical turbulence and its applications to the topic of primordial magnetic fields. In section \ref{se:numerical_set_up}, we give basic definitions and describe the numerical set-up. In \ref{se:p_q_time_evolution} we discuss the subtleties involved in the measurement of the decaying exponents. In section \ref{se:results_HD_vs_MHD}, we make a comparison between a hydrodynamic, a helical and a nonhelical MHD simulation. In section \ref{se:decay_at_varying_Pm}, we perform a detailed analysis of decaying helical and nonhelical MHD for varying Prandtl number. In section \ref{se:varying_Re} we do the same for varying Reynolds number, and in section \ref{se:varying_kp} for the scale separation. In section \ref{se:hyperviscous}, we show the results of a small number of simulations that use hyperviscosity and hyperresistivity, to study the effect that these have on the inverse transfer. We dedicate section \ref{se:comparison_literature} to the comparison between our results and those produced by other codes in literature. 

\section{Decaying turbulence}
\label{se:decaying_turbulence}

\subsection{Decaying hydrodynamics}

The first to establish the laws of decaying turbulence was Kolmogorov in one of his foundational works in 1941 \cite{kolmogorov1941degeneration}. This theory is based on the approximate invariance of the Loitsyansky integral $I = - \int d\bm{r} \, \bm{r}^2 \langle \bm{u}(\bm{x}) \bm{u}(\bm{x}+\bm{r})\rangle \sim L^5 U^2 = 2 L^5 E$, where $L$ is the integral length scale, $E$ the energy, and $U$ the root mean squared velocity. This integral is used to set the timescale of the energy decay, assuming that $dE/dt \sim - E/T \sim - E^{3/2}/L \sim - E^{17/10}$ the energy decay rate follows the power law
\begin{equation}
    E  \propto t^{-10/7}.
\end{equation}

The invariance of the Loitsyansky integral has been challenged by some authors years later. These suggest that long-range correlation may exist in turbulent flows depending on the form of the initial spectrum ($\sim k^2$ or $\sim k^4$) \cite{batchelor1953theory,batchelor1956large,birkhoff1954fourier,saffman1967large,davidson1997role,davidson2000loitsyansky}. This gives place to multiple predictions for the decaying laws, that depend on initial conditions and other physical assumptions. For instance, in 1967, Saffman predicted a decay law of the form $E \sim t^{-6/5}$ for an initial spectrum of the form $E(k)\sim k^2$ \cite{saffman1967large}. 

Furthermore, numerical and experimental results add to the controversy by showing a variety of results for the decaying exponent $p$ (i.e. $E\sim t^{-p}$), as well as different conclusions for the dependence of $p$ on initial conditions and experiment/simulation set-up. For experimental results see \cite{comte1966use,comte1971simple,mohamed1990decay,skrbek2000decay,thormann2014decay,sinhuber2015decay} and for numerical results see \cite{ishida2006decay,yoffe2018onset,meldi2017turbulence,mansour1994decay,yu2006direct,meldi2011stochastic,thornber2007implicit,anas2020freely,huang1994power}. There are many factors that might affect the evolution of the decay in either numerical simulations or experiments. For instance, even though the theory assumes the Re$\rightarrow\infty$ limit, numerical and experimental measurements are done at finite Reynolds numbers, preventing the formation of the self-similar cascade on which the theories rely. Furthermore, some authors notice the importance of large scale resolution, suggesting that a saturation occurs when the size of the largest eddies becomes similar to size of the box, thus affecting the dynamics of the decay  \cite{skrbek2000decay,thornber2016impact,meldi2017turbulence}. 

These controversies brought recent attention to the problem of obtaining truly universal laws for decaying turbulence \cite{yoffe2018onset,panickacheril2022laws}. In \cite{panickacheril2022laws}, the authors collected data from different numerical and experimental results reported over decades, which was analyzed to provide some clarity on this issue.

\subsection{Decaying MHD}

A topic of major theoretical and practical importance, is the decay of MHD turbulence. In MHD, the difficulty to establish universal decaying laws goes one step further. The variety of physical situations and initial conditions is much wider than in hydrodynamics, and the interplay of time and length scales is much more complex. Still, multiple predictions can be made based on different assumptions. 

A common classification for MHD flows is done in terms of the value of the magnetic helicity, which is defined as

\begin{equation}
    \langle H_b \rangle = \lim_{V\to\infty}V^{-1} \int_V dV\, \langle \bm{A} \cdot \bm{B}\rangle \quad ,
    \label{eq:Helicity_conserv}
\end{equation}
where $\bm{B}$ is the magnetic field, $\bm{A}$ is the magnetic vector potential such that $\nabla \times \bm{A} = \bm{B}$. The magnetic helicity is an ideal invariant in MHD, and its turbulent dynamics present interesting properties. The pioneering paper by Frisch et al. \cite{frisch1975possibility} showed that the magnetic helicity spectrum $H_b(k,t) = \int_{\lvert\bm{k}\rvert= k} \langle\bm{A}(\bm{k})\cdot\bm{B}(-\bm{k})\rangle$ exhibits an inverse cascade that had not been previously predicted or observed for kinetic helicity in pure hydrodynamic flows. Additionally, in this paper it is shown that the flow must obey the following realizabilty condition

\begin{equation}
\lvert H_b(k,t)\rvert\leq \frac{2 \lvert E_b(k,t)\rvert}{k} \quad , 
    \label{eq:realizability_condition}
\end{equation}
where $E_b(k,t)$ is the magnetic energy spectrum. This suggests that the inverse cascade of magnetic helicity might be accompanied by a transfer of magnetic energy to large scales in order to satisfy the condition in Eq. (\ref{eq:realizability_condition}). The implication of this process is that magnetic energy is redistributed to larger scales, leading to the creation of coherent magnetic structures at scales much larger than the ones where the energy was initially injected. Over the decades, numerous numerical evidence have supported the existence of such an inverse cascade \cite{alexakis2006inverse,balsara1999formation,berera2014magnetic,brandenburg2001inverse,muller2012inverse,pouquet1976strong}. 

We refer to flows as \textit{helical} if the Eq. (\ref{eq:realizability_condition}) satisfies the equality for a most values of $k$. In this case, it is said that magnetic helicity is maximal. Alternatively, we define the flow to be \textit{nonhelical} if the magnetic helicity is practically null. In the nonhelical case, the mechanisms and phenomenology described above are not present in nonhelical flows. Apart from these two cases, MHD flows can also be partially helical, but in this work, we only focus on cases where magnetic helicity is either maximal or vanishing.

For decaying turbulent MHD flows, the difference between helical and nonhelical flows is also remarkable. In the helical case, magnetic helicity is approximately conserved during the decay, even for non-vanishing resistivity, hence, looking at Eq. (\ref{eq:Helicity_conserv}) it can be estimated that

\begin{equation}
    \langle H_b \rangle  \sim B(t)^2 L(t) \sim E_b(t) L(t)\quad ,
    \label{eq:Helicity_conserv2}
\end{equation}

where $B$ is the rms magnetic field and $L$ is the coherence length of the field.

This conservation results in an inverse cascade of magnetic helicity from small to large length scales, that supports an inverse transfer of magnetic energy. This also produces a slower energy decay and a faster growth of the integral lengthscale than in the hydrodynamic case. 

For $U \sim B$ and $L_b \sim L_u$, it can be derived that the magnetic and kinetic energy decay with similar rates $E_b \sim E_u \sim t^{-2/3}$ \cite{hosking2021reconnection}. Nevertheless, the magnetic field decay measured in numerical simulations is shallower than this prediction \cite{banerjee2004evolution,biskamp1999decay,christensson2001inverse,berera2014magnetic,brandenburg2017classes}. 

In \cite{hosking2021reconnection,zhou2019magnetic,zhou2020multi,zhou2021statistical,zhou2022scaling}, it is proposed that magnetic reconnection is the mechanism leading the inverse transfer and setting the decay timescale. Based on these arguments, in \cite{hosking2021reconnection} different decay rates are computed depending on the situation. For $U\ll B$, it is found that $E_b \sim t^{-4/7}$ and $E_u \sim t^{-5/7}$, whereas for $U\sim B$, it is conjectured that an invariant related to the cross-helicity (even in cases with no net cross-helicity), produces a decay of the form $E_u\sim t^{-1}$ and $E_b\sim t^{-1/2}$. 

In the nonhelical case, the conservation of magnetic helicity cannot be used to estimate the timescale of the decay. The nonhelical inverse transfer found recently in numerical work has brought increasing interest in this topic. A decade ago, multiple works reported this inverse energy transfer using DNS \cite{brandenburg2015nonhelical,berera2014magnetic,zrake2014inverse}, although many years before, in \cite{christensson2001inverse} they had observed a small inverse transfer of energy in direct numerical simulations for the nonhelical case, suggesting that this effect could be more pronounced at larger values of Re. The physical mechanisms behind this inverse transfer remain unclear and recent studies were dedicated to this topic \cite{linkmann2016helical,park2017inverse,reppin2017nonhelical,bhat2021inverse,hosking2021reconnection}.    

Campanelli analyzed the problem of decaying nonhelical MHD using Olesen scaling arguments. These explore the self-similarity properties of MHD equations \cite{olesen1997inverse}, reaching to the conclusion that for an initial spectra of the form ($E_b(k,t) \sim E_u(k,t) \sim k^{4}$), the magnetic energy decays as $E_b \propto t^{-1}$  \cite{campanelli2014evolution,campanelli2007evolution,campanelli2004scaling}.

In \cite{brandenburg2015nonhelical}, the \texttt{PENCIL} code is used and a clear inverse transfer of magnetic energy is observed for a magnetically dominated case. The authors suggest that the inverse transfer occurs either due to the shallower kinetic spectra $\sim k^2$ that dominates at large scales, interacting with the magnetic field to force larger coherence scales, or due to the local two-dimensional behavior of the magnetic vector potential that might enhance the inverse transfer. In this work, a weak turbulence spectrum ($\sim k^{-2}$) is observed in the inertial range. Runs that start with small helicity show an energy decay that goes like $t^{-1}$ initially, and then approach a $t^{-1/2}$ as the system approaches maximal helicity. The $t^{-1}$ scaling has been repeatedly observed in numerical simulations of nonhelical MHD \cite{biskamp1999decay,berera2014magnetic,christensson2001inverse,tevzadze2012magnetic,reppin2017nonhelical,zrake2014inverse,bhat2021inverse}.
In \cite{berera2014magnetic,zrake2014inverse}, an inverse energy transfer and a magnetic energy decay of $t^{-1}$ are also reported.

Recently in \cite{bhat2021inverse}, magnetic reconnection is proposed as the main mechanism for the inverse transfer, and this is investigated using DNS of initially magnetically dominated flows as well as a kinetic dominated case. The authors show that if the time is normalized using the magnetic reconnection timescale, all decay curves with different Lundquist numbers collapse into each other. When the flow is initialized with non-zero kinetic energy, a weaker but present inverse transfer is found, possibly related to the dynamo action at large scales.

In \cite{hosking2021reconnection}, the authors give an intuitive explanation of the mechanism leading the decaying timescale based on small positive and negative helical structures. Similarly to the conservation of the Loitsyansky integral in hydrodynamics, the authors propose that the finite contributions from these helical structures conserve the integral $I_H = \int d\bm{r} \langle h(\bm{x}h(\bm{x}))\rangle$, where $h=\bm{A}\cdot\bm{B}$ is the helicity density. This results in a conservation law of the form 

\begin{equation}
    B^4 L^5 \sim E_b^2 L^5 \sim \text{const} \quad .
\end{equation} 

For the magnetically dominated case, this gives a magnetic energy decay of the form $t^{-1.18}$ in the slow reconnection regime and $t^{-1.11}$ in the fast reconnection regime. Nevertheless, for the case in which $U\sim B$, it is conjectured that a Saffman-type invariant associated to cross-helicity gives a decay of $t^{-10/7}$ for both the magnetic and kinetic energy, same as the Loitsyanksy-Kolmogorov prediction for hydrodynamic flows. 

Most of the decaying nonhelical MHD studies found in literature have been done for Prandtl number Pm $=1$, except from \cite{reppin2017nonhelical,brandenburg2015nonhelical,mckay2019fully,park2017inverse}. In \cite{reppin2017nonhelical}, a thorough numerical study using the \texttt{PENCIL} code is performed, with a wide parameter range exploration varying Pm and scale separation. These simulations show that the inverse transfer is suppressed for increasing Prandtl number. The authors argue that this occurs due to the slow magnetic reconnection at high Pm. The opposite behavior is observed in \cite{mckay2019fully}, where the growth of magnetic energy at large scales grows for increasing Pm.

\subsection{Application to the decay of primordial magnetic fields}

The prevalence of cosmic magnetic fields at different locations and scales in space presents a unique link to the physics of the early universe through present--day observations, particularly (but not exclusively) the cosmic microwave background. The presence of cosmic fields in voids of the large scale structure, with strengths of $10^{-16}$ G and coherence lengths of Mpc scales, are thought to be of primordial origin \cite{Subramanian:2015lua, Shukurov2021}.

A major subject to explain the strength and scale of the observed magnetic fields, is the evolution of the magnetic fields across the multiple cosmological eras. This goes back to the seminal paper of Turner and Widrow \cite{Turner:1987bw}, where the generation of cosmic magnetic fields in the context of inflation was explored. Next, Brandenburg, Enqvist and Olesen found covariant MHD equations for an expanding spacetime \cite{brandenburg1996large,brandenburg1997effect}. Finally, the first comprehensible numerical and analytical study of the evolution of magnetic fields throughout the (standard) cosmological expansion was given by Banerjee and Jedamzik \cite{banerjee2004evolution}. 

Accounting for the expanding background introduces new terms to the usual MHD equations. This is hardly unexpected, as even for the simple evolution of a ``free'' magnetic field, its strength varies as $1/a^{2}$ --- with $a(t)$ denoting the scale factor --- due to flux conservation. However, it was pointed out in \cite{brandenburg1996large, Subramanian:1997gi} that for the radiation--dominated era, a convenient rescaling of the MHD variables can render the same set of equations as in flat spacetime. In this way, one can use standard analytical and numerical tools to perform the MHD analysis in a cosmological context. For the matter--dominated era, one can introduce a different set of variables that can render similar equations to those of standard MHD \cite{Martel:1997hk,Doumler:2009vq}. Even though expansion--related terms persist, which effectively slow down the rate of dynamical evolution, standard MHD equations can be applicable for periods of time where the typical microphysical processes are faster than the expansion rate, or equivalently, where the corresponding characteristic times are smaller than one expansion time. On the other hand, it has also been reported that the assumption of incompressibility is valid during the radiation-dominated era due to the large value of the speed of sound in a relativistic plasma. This could be extended to the matter-dominated era, except for field strengths leading to current values larger than $10^{-11}$G, where the fluid is not compressible after photon decoupling. Thus, as long as one stays below that value, the results we obtain in nonrelativistic and incompressible MHD could be, in principle, applied to this cosmological period. 

Most numerical works are done at moderate Prandtl numbers. However, the interstellar and intergalactic medium have an estimated Prandtl number in the range $10^8-10^{14}$ \cite{schober2012magnetic,verma2004statistical,plunian2013shell,federrath2014turbulent,banerjee2004evolution,kulsrud1999critical,schekochihin2002small}. Even though these numbers are not practical for DNS, we can study the trends in the behavior for increasing Pm. This can give some hints to understand a more realistic scenario.

\section{Numerical set-up}
\label{se:numerical_set_up}

We investigate the decay of MHD turbulence using fully resolved DNS for a wide range of parameters.
For this we use the EddyBurgh code \cite{yoffe2013investigation,linkmann2016self}. This solves the incompressible MHD equations given by 
\begin{subequations}
\begin{align}
    \partial_t \bm{u} &= - \left(\bm{u} \cdot \nabla \right) \bm{u} + \nu \nabla^2 \bm{u} + \left(\nabla \times \bm{B} \right) \times \bm{B} ,  \label{eq:MHDeqnsu} \\
    \partial_t \bm{B} &= \left( \bm{B}\cdot\nabla\right) \bm{u} - \left(\bm{u} \cdot \nabla\right) \bm{B} + \eta \nabla^2 \bm{B} \quad , 
    \label{eq:MHDeqnsb}
\end{align}
\label{eq:MHD_Eqns}
\end{subequations}
where $\nu$ is the viscosity and $\eta$ the resistivity. We use Alfvenic units, so the mass density $\rho=1$ and the Alfv\'en velocity $v_A = B$.

EddyBurgh is a pseudospectral DNS code that solves MHD equations (\ref{eq:MHD_Eqns}) on a cubic box of length $\ell_{\text{box}}=2\pi$ with periodic boundary conditions. This consists of a cubic lattice of $N^3$ equally spaced collocation points, where $N = 256,512,1024,2048$ and $4096$. The $2/3$ dealiasing rule is implemented. We use a predictor-corrector time-stepping procedure, also known as Heun's method \cite{heun1900neue}. 

To initialize the fields in a given helical state, we use a orthonormal helical basis $\bm{h}_{\bm{k}}^{+}$, $\bm{h}_{k}^{-}$, which are eigenvectors of the curl operator, so they satisfy $i\bm{k}\times\bm{h}_{\bm{k}}^{+} = k\bm{h}_{\bm{k}}^{+}$ and $i\bm{k}\times\bm{h}_{\bm{k}}^{-} = -k\bm{h}_{\bm{k}}^{-}$. These are fully helical by construction. Furthermore, given that the fields are solenoidal, we can expand the three-dimensional kinetic and magnetic fields into this helical basis, i.e. $B(\bm{k},t) = B^{+}(\bm{k},t)\bm{h}_{\bm{k}}^{-} + B^{-}(\bm{k},t)\bm{h}_{\bm{k}}^{+}$ for the magnetic field and likewise for the kinetic field. For further details, we refer the reader to Refs. \cite{constantin1988beltrami,lessinnes2009helical,waleffe1992nature}. Our runs are initialized either with maximal magnetic helicity (labeled with H) or with vanishing helicity (labeled with NH). For this we first initialize the fields by assigning a Gaussian-random vector to the fields at each point $\bm{x}$. Then, we apply a Fourier transform and we expand the resulting fields onto the helical basis to set the desired initial helical state. Setting the initial state as nonhelical is rather straightforward, as the Gaussian-random vectors gives an distribution of positive and negative helical modes that cancel each other out, resulting in a state with no net helicity. In order to set the initial state as fully helical, we keep only one of the helical projections $B^{+}$ or $B^{-}$ (the choice is indistinct). In this work, kinetic and cross-helicities are set to zero initially. Finally, we rescale the fields to obtain the desired initial spectra. All our runs have an initial magnetic spectrum of the form
\begin{gather}
    E_b(k) = E_b(t=0) \,  c_1 \left(\frac{k}{k_p}\right)^{4} \exp\Bigg[-2 \left(\frac{k}{k_p}\right)^{2}\Bigg]  \quad ,  \\
    c_1 = \frac{2^{11/2}}{3\sqrt{\pi}}\, k_p^{-1} \quad  , 
\end{gather}
where $k_p$ is the wavenumber at which the spectrum peaks, and $c_1$ is a normalisation factor that ensures that the initial energy remains invariant under changes of $k_p$. This spectrum, has a form of $\sim k^{4}$ for $k<k_p$, it peaks at $k=k_p$ and shows a sudden cut-off for $k>k_p$.

All simulations are freely evolving from $t=0$. Most runs are initially in equipartition (i.e. $E_{u}(k,t=0) = E_{b}(k,t=0)$), although we run a smaller number of simulations that are initialized with the velocity field set to zero. The integral (or coherence) lengthscales of the kinetic and magnetic field are computed as $L_{u,b} = \left(3\pi/4 E\right) \int_{0}^{\infty} dk\, E_{u,b}(k)/k$. The rms velocity is defined so that $E_u=3 U^2/2$ and the rms magnetic field such that $E_b = 3B^2/2$. The large eddy turnover time is $T_u= L_u/U$ and the Alfv\'en time is $T_b = L_b/B$. We use the initial eddy turnover time as a reference. Runs with equipartition satisfy $U(0)=B(0)$ and $L_u(0)=L_b(0)$, hence, we refer to the large eddy turnover time simply as $T = T_u(0) = T_b(0)$. In the magnetically dominated case, $T=T_b(0)$, as $T_u$ is not defined for $t=0$. In the same way we express the initial Reynolds number Re$(0)=U(0)L(0)/\nu$ as Re, note that this is not defined when the velocity field is initialized to zero. We also refer to the initial Lundquist number $S(0)=B(0)L_b(0)/\eta$ as $S$. The magnetic Reynolds number Re$_b =UL/\eta$ and the Lundquist number are equivalent at $t=0$ for all runs initialized in equipartition. The kinetic and magnetic dissipation rates are computed as $\varepsilon_{u}= 2\nu \int dk\,E_u k^{\gamma}$ and $\varepsilon_b = 2\eta \int dk \, E_b(k) k^{\gamma}$, where $\gamma=2$ for standard viscosity and $\gamma=4$ for the hyperviscous/hyperresistive case. We keep all simulations fully resolved unless otherwise stated. We consider runs fully resolved when both the kinetic dissipative scale $k_{\nu} = \left(\varepsilon_u(t)/\nu^3\right)^{1/4}$ and the magnetic dissipative scale $k_{\eta} = \left(\varepsilon_b(t)/\eta^3\right)^{1/4}$ satisfy $k_{\text{max}}/k_{\nu}>1.25$ and $k_{\text{max}}/k_{\nu}>1.25$ throughout the entire duration of the run. Only in one extreme case we get $k_{\text{max}}/k_{\eta} \approx 1.15$. 

We run two cases using hyperviscosity and hyperresistivity, with the aim of establishing a qualitative comparison with previous results in literature. This consists in modifying the viscous and resistive terms in Eqs. (\ref{eq:MHDeqnsu}) and (\ref{eq:MHDeqnsb}) respectively, so the gradient is now of fourth order, i.e. $\nu\nabla^2 \rightarrow \nu_2\nabla^4$ and $\eta\nabla^2 \rightarrow \eta_2 \nabla^4$.
In all nonhelical runs, $H_b/2E_bL_b \sim \mathcal{O}(10^{-3})$. This ensures that the simulations remain practically nonhelical for all times.

The aim of this work is to perform a wide parameter study of MHD decay by measuring the spectra evolution and the scaling laws for a number of high resolution numerical simulations of helical and nonhelical MHD turbulence. We pay special attention to the inverse transfer of energy.

\section{Scaling exponents $p$ and $q$}
\label{se:p_q_time_evolution}

In this work we report scaling exponents for the kinetic and magnetic energy decay $E_{u,b}\sim t^{-p_{u,b}}$, and coherence length growth $L_{u,b} \sim t^{q_{u,b}}$. Measuring these exponents is not always a straightforward task. The first feature of the scaling exponents we find in our simulation is that they are not constant throughout the decay in all cases. To see this, we measure
\begin{subequations}
\begin{align}
p_b(t) &= - \frac{d \log E_b}{d \log t} \quad ,\\
q_b(t) &= - \frac{d \log L_b}{d \log t} \quad .
\end{align}
\end{subequations}

 Typically, there is an initial transient of a few eddy turnover times for the system to reach an approximate self-similar decay (see section \ref{se:results_HD_vs_MHD}, Figures \ref{fig:HD_HEL_NONHEL_spectra} and \ref{fig:HD_HEL_NONHEL_scaling_exponents}). When Re is high, the system decays while turbulence is developed and the decay follows a power law $t^{-p}$. During this stage, a plateau is observed in the evolution of $p(t)$. Nevertheless, in some other cases, for moderate and low Re, there is no clear time interval in which $p$ and $q$ reach a steady constant value. This introduces certain ambiguity in the measurements. For very low Reynolds numbers, turbulence is not developed at all and the decay is almost entirely diffusive. Note that this becomes slightly more difficult when we vary the Prandtl number, for instance, for a low Pm, we can have a turbulent kinetic decay together with a diffusive magnetic decay, only kept up by dynamo effect.  

We also note that when the scale separation $k_p \sim \mathcal{O}(1)$, the initial transient tends to be slower, not only taking more time for the energy to reach a power-law decay but also adding a larger bias, since the approximation $(t-t_0)^{-p} \approx t^{-p}$ becomes more questionable, without clear arguments to determine $t_0$ as pointed out in \cite{biskamp1999decay}. Furthermore, for low Re, diffusion takes over rather quickly, not allowing the development of turbulence at $t \gg t_0$. 

We measure the scaling exponents by determining a time interval in which an approximate plateau is observed in the evolution of $p(t)$ and $q(t)$. Then we perform a linear regression with logarithmic axes within that time interval. In some cases, when we compare multiple cases, there is no clear common plateau in the evolution of their exponents. In those cases we show the time evolution $p(t)$ and $q(t)$ and discuss the criteria used. We acknowledge the bias that is introduced by using logarithmic axes. An alternative method to overcome these problems is used in \cite{hosking2021reconnection} to make comparisons to predictions, although this is also subject to the subjective choice of intervals. 

Nevertheless, the main goal of this work is not to contrast our measurements against precise theoretical predictions. Instead, we look at trends as we vary different parameters and see whether our measurements are consistent with these predictions or not. In fact, discriminating between different theoretical predictions that are very close, requires a careful numerical treatment that is often ignored in literature.  

In order to obtain a fully turbulent decay, we need large Reynolds numbers. This requires higher resolutions as we need to resolve increasingly large wavenumbers $k_{\nu}$ and $k_{\eta}$. However, if we want to observe the build-up of magnetic energy at large scales and its influence in the decay, we need to have enough scale separation, i.e. $k_p \gg 1$, but this puts more energy closer to the dissipation scales, which prevent us from obtaining a large Reynolds number for a given resolution. This is especially challenging when it comes to explore cases with Pm $ \gg 1$ at the same time we keep a relatively high Re. For low Re, the smaller scales are expected to decay as $\exp\left(-2 \nu k^2 t\right)$, if most of the energy decays through viscosity (or diffusivity in the case of magnetic energy), the decay is approximately of the form $E\sim t^{-5/2}$ \cite{batchelor1948decay,mansour1994decay}. In section \ref{se:Helicity conservation} we provide more arguments that show that slight details can significantly affect the predicted values of $p$ and $q$ at finite resistivity.

Some authors use hyperviscosity and hyperresistivity to overcome resolution limitations \cite{reppin2017nonhelical,hosking2021reconnection}. This allows a larger inertial range without a greater demand on resolution. The cost is that this alters the theoretical predictions, as it introduces different scaling relations that we prefer to avoid in this analysis and also because modifying the resistive region affects the magnetic reconnection process. Still, we run a few hyperviscous simulations just for the sake of qualitative comparison with other results in literature.

\section{MHD vs. HD decay}
\label{se:results_HD_vs_MHD}

First we show the spectral evolution and the scaling laws of the kinetic energy decay of a helical and a nonhelical flow with Pm $=1$. We compare these to a pure hydrodynamic decay with the same viscosity $\nu=0.0003125$, $k_p = 40$ and $N = 2048$. Both MHD simulations are initially in equipartition so the kinetic spectra are equivalent for all three runs at time $t=0$.

\begin{figure}
    \centering
    \includegraphics[width=\linewidth]{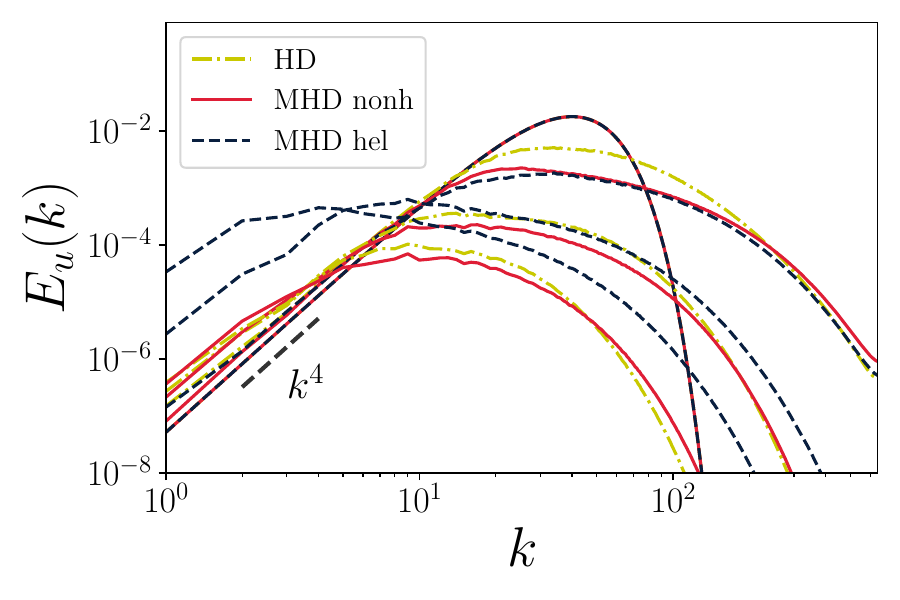}
    \caption{Kinetic energy spectra during decay for runs H\textsubscript{Re5}, NH\textsubscript{Re5}, and a hydrodynamic run with same parameters and same initial spectrum. Different curves correspond to times $t/T = 0$, $1.6$, $12.8$ and $50$.}
    \label{fig:HD_HEL_NONHEL_spectra}
\end{figure}

In Figure \ref{fig:HD_HEL_NONHEL_spectra}, we see that the helical case shows a strong inverse transfer of energy in the evolution of the kinetic spectra consequence of the approximate conservation of the mean magnetic helicity in the helical case. Instead, the nonhelical case shows only a slight increase of kinetic energy at large scales than the hydrodynamic case, which shows that even for the highest Reynolds number we achieve, the inverse transfer of kinetic energy is not significant. 

\begin{figure}
    \centering
    \includegraphics[width=\linewidth]{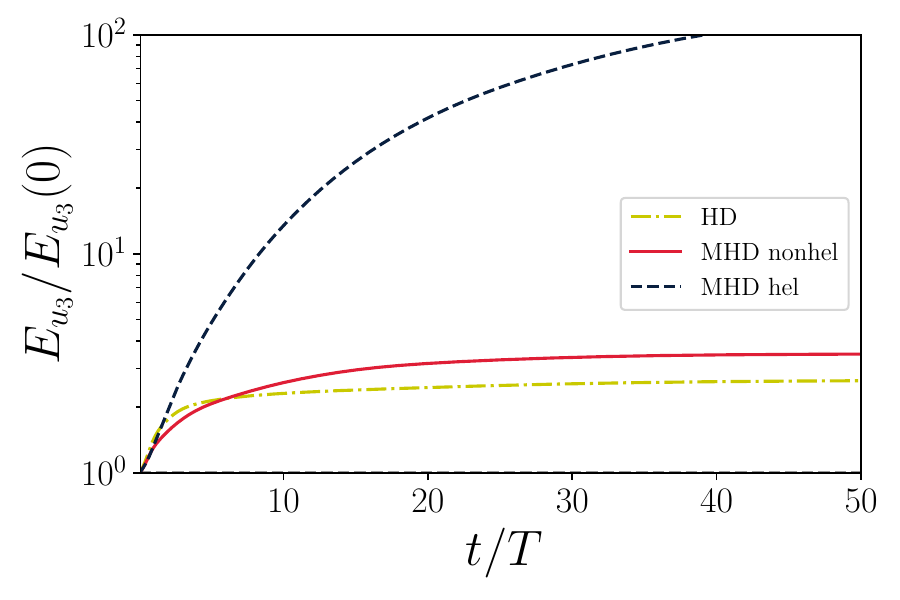}
    \caption{Time evolution of large scale kinetic energy $E_{u_3}(t)/E_{u_3}(0)$ for runs H\textsubscript{Re5}, NH\textsubscript{Re5}, and a hydrodynamic run with same parameters and same initialization.}
     \label{fig:HD_HEL_NONHEL_E3_vs_t_comparison}
\end{figure}

This comparison becomes clear when we observe the energy at large scales as shown in Figure \ref{fig:HD_HEL_NONHEL_E3_vs_t_comparison}, where we measure the growth of energy for $k \leq 3$ as $E_{u_3} = \int_{0}^3 dk \, E_u(k) $. In any case, the growth of energy at large scales is much weaker than the one found in \cite{brandenburg2017classes}, where the effect is much more pronounced. In principle, this small difference between the hydrodynamic and the nonhelical case can be either due to a stronger inverse cascade of kinetic energy in MHD, or due to the transfer from magnetic to kinetic energy at large scales due to the local action of the Lorentz force. However, we cannot discard that this difference is merely related to box size effects.

It is clear that the helical run shows a larger growth of coherence length than both the nonhelical and the pure hydrodynamic case. This can be seen from the evolution of the spectrum peak, that is a fair estimate of the coherence length evolution.

\begin{figure}
    \centering
\subfloat[$p_u(t)$]{
    \includegraphics[width=0.5\linewidth]{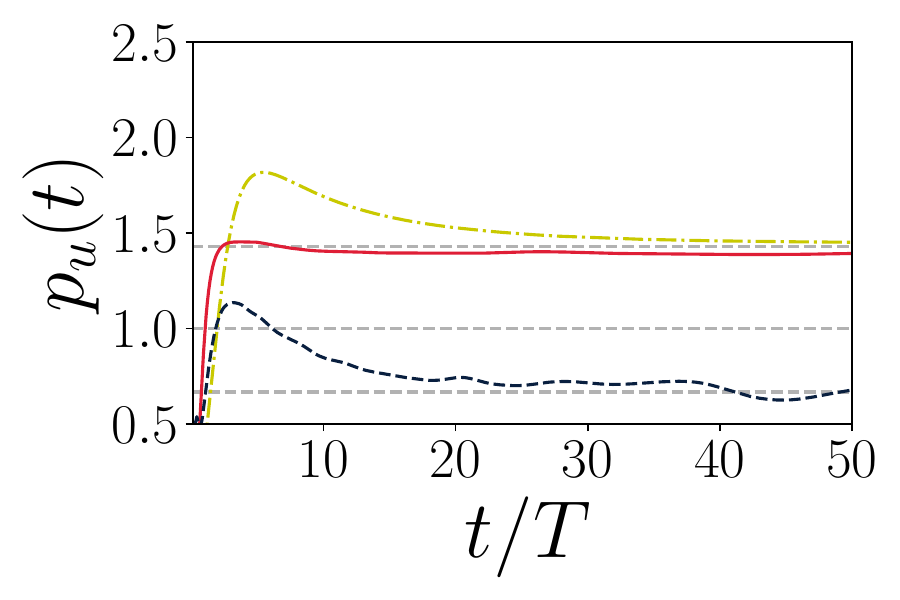}
    \label{fig:HD_HEL_NONHEL_p_t}
}
\subfloat[$q_u(t)$]{
    \includegraphics[width=0.5\linewidth]{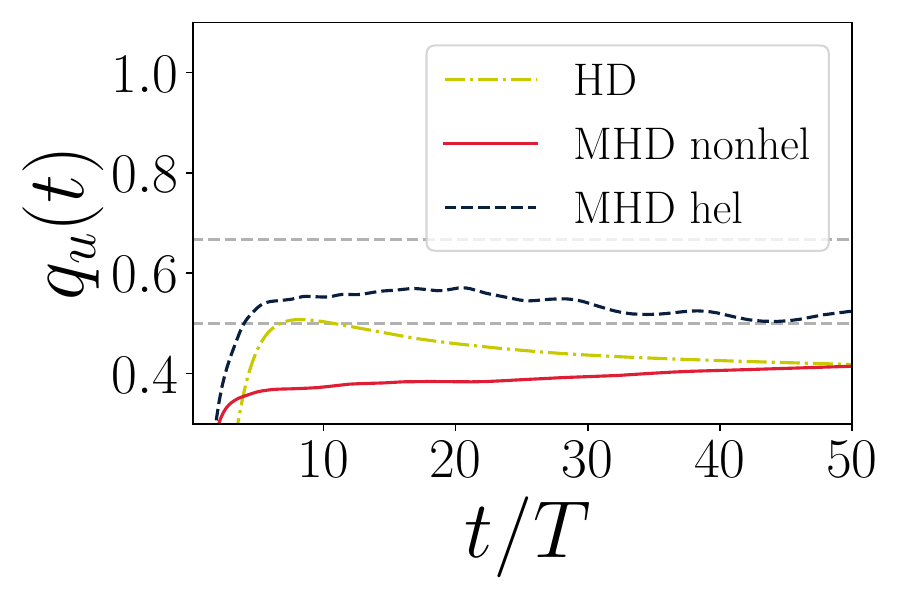}
    \label{fig:HD_HEL_NONHEL_q_t}
}
     \caption{Time evolution of the scaling exponents $p_u$ (a) and $q_u$ (b) for runs H\textsubscript{Re5}, NH\textsubscript{Re5}, and a hydrodynamic run with same parameters and same initial spectrum. The horizontal gray dashed lines correspond to typical values $p_u = 1$, $10/7$ and $2/3$, and $q_u = 1/2$ and $2/3$. }
     \label{fig:HD_HEL_NONHEL_scaling_exponents}
\end{figure}

Finally, we look at the time evolution of the scaling exponents $p_u$ and $q_u$ in Figure \ref{fig:HD_HEL_NONHEL_scaling_exponents}. We see that the hydrodynamic and the nonhelical MHD run have different transients but end up decaying at the same rate, consistent with the Loitsyansky-Kolmogorov prediction $t^{-10/7}$. The integral scale grows approximately like $t^{0.4}$ in both cases. On the other hand, the helical flow show different properties, with a shallower decay that goes like $t^{-2/3}$ and the integral scale grows approximately like $t^{1/2}$. Interestingly, the helical exponents present a more erratic behavior, possibly due to the integral lengthscale becoming comparable to the box size much faster than in the other cases.

\section{Decay at varying Prandtl number}
\label{se:decay_at_varying_Pm}

We now look at the decay of helical and nonhelical MHD for varying Pm. 
Most works in literature have been carried out for a short range of values for Pm, except from \cite{reppin2017nonhelical} and \cite{mckay2019fully}. In \cite{reppin2017nonhelical}, a thorough analysis at varying Prandtl number is done using both standard viscosity and hyperviscosity. The authors find that as Pm increases, the growth of magnetic energy at large scales is suppressed. In our simulations we find the opposite trend. We comment on this in section \ref{se:comparison_literature}. 

We run a set of 22 simulations initialized in equipartition, 11 helical and 11 nonhelical (see Table \ref{tab:varying_Pm_v0.005_kp_40}). The runs with Pm $=32$ is slightly underresolved, with $k_{\text{max}}/k_{\eta}\approx 1.15$ during the first turnover times. We also run 8 simulations with the velocity field initialized to zero, 4 helical and 4 nonhelical (see Table \ref{tab:varying_Pm_v0.005_kp_40_zero}). In every case the scale separation is set at $k_p=40$ and viscosity is fixed at $\nu = 0.005$. This choice of parameters results in a moderate Reynolds number, preventing the kinetic flow from developing a complete turbulent state. Still, we increase Prandtl number while keeping constant viscosity and we measure the scaling exponents at $t \approx 50T$. The reason for this is that $p(t)$ and $q(t)$ do not reach a constant value for higher Pm. Instead, they tend to an asymptotic value as time evolves. For this reason, we consider the value of $p(t)$ and $q(t)$ at the latest time measured ($t= 50T$) as the best estimate of the exponent.

\subsection{Scaling exponents for initial equipartition}
\label{se:v_0.005_Pm}

\begin{figure*}[ht]
\centering
\subfloat[$p$ vs. Pm]{
  \includegraphics[width=0.5\linewidth]{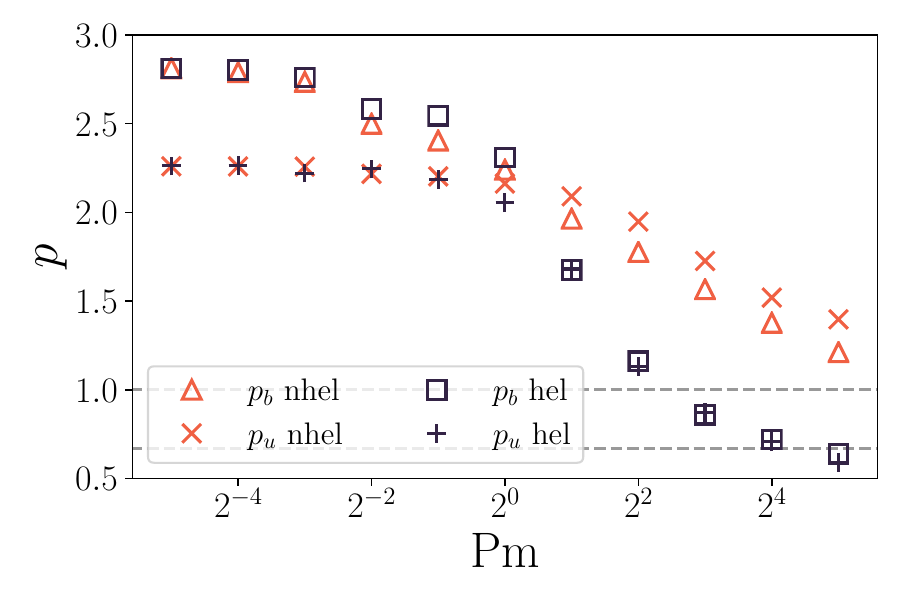}
  \label{fig:p_vs_Pm_v0.005}
}
\subfloat[$q$ vs. Pm]{
  \includegraphics[width=0.5\linewidth]{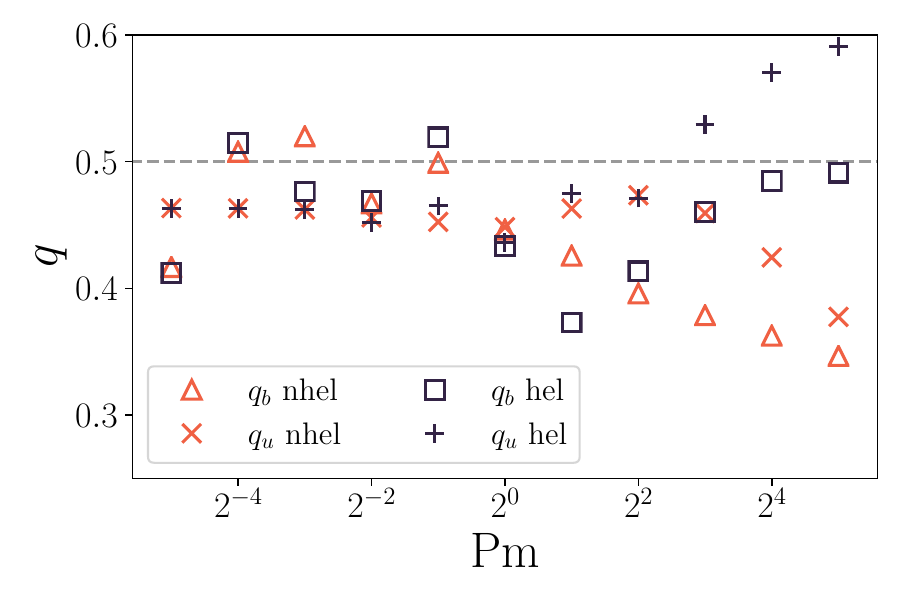}
  \label{fig:q_vs_Pm_v0.005}
}
\caption{Scaling exponents $p$  (a) and $q$ (b) for the following cases: magnetic helical (blue squares), magnetic nonhelical (orange triangles), kinetic helical (blue crosses) and kinetic nonhelical (orange crosses). Dashed horizontal lines correspond to the typical scaling values observed in literature $p=1$ for nonhelical flows and $q=2/3$ for helical flows. All runs are initially in equipartition.}
\label{fig:p_and_q_v0.005}
\end{figure*}

First, we analyze the case in which runs are initially in equipartition. In Figure \ref{fig:p_vs_Pm_v0.005} we show the kinetic and magnetic scaling exponents for varying Pm, whereas in Figure \ref{fig:q_vs_Pm_v0.005} we show the scaling exponents measured for the magnetic coherence scale.

\begin{table*}[ht]
    \centering
\begin{tabular}{llrrrr}
\toprule
\toprule
   Run &     Pm &  $\nu$ &  Re &  $k_p$ &   $N$ \\
\midrule
 H/NH\textsubscript{p-5} &   $2^{-5}$ &  0.005 &     8 &     40 &   512 \\
 H/NH\textsubscript{p-4} &   $2^{-4}$ &  0.005 &     8 &     40 &   512 \\
 H/NH\textsubscript{p-3} &   $2^{-3}$ &  0.005 &     8 &     40 &   512 \\
 H/NH\textsubscript{p-2} &   $2^{-2}$ &  0.005 &     8 &     40 &   512 \\
 H/NH\textsubscript{p-1} &   $2^{-1}$ &  0.005 &     8 &     40 &   512 \\
 H/NH\textsubscript{p0} &   1 &  0.005 &     8 &     40 &   512 \\
 H/NH\textsubscript{p1} &   2 &  0.005 &     8 &     40 &  1024 \\
 H/NH\textsubscript{p2} &   $2^{2}$ &  0.005 &     8 &     40 &  1024 \\
 H/NH\textsubscript{p3} &   $2^{3}$ &  0.005 &     8 &     40 &  2048 \\
 H/NH\textsubscript{p4} &  $2^{4}$ &  0.005 &     8 &     40 &  2048 \\
 H/NH\textsubscript{p5} &  $2^{5}$ &  0.005 &     8 &     40 &  2048 \\
\bottomrule
\bottomrule
\end{tabular}
    \caption{Helical and nonhelical runs for varying Prandtl number and fixed viscosity $\nu = 0.005$ and $k_p = 40$. All runs are initially in equipartition.}
    \label{tab:varying_Pm_v0.005_kp_40}
\end{table*}

\begin{table*}[ht]
    \centering
    \begin{tabular}{lrrrrr}
    \toprule
    \toprule
       Run &     Pm &  $\nu$ &   $S$ &  $k_p$ &   $N$ \\
    \midrule
     H/NH\textsubscript{pz-2} &   0.25 &  0.005 &    2 &     40 &   512 \\
     H/NH\textsubscript{pz0} &   1.00 &  0.005 &    8 &     40 &   512 \\
     H/NH\textsubscript{pz2} &   4.00 &  0.005 &   32 &     40 &  1024 \\
     H/NH\textsubscript{pz4} &  16.00 &  0.005 &  129 &     40 &  2048 \\
    \bottomrule
    \bottomrule
    \end{tabular}
    \caption{Helical and nonhelical runs for varying Prandtl number and fixed viscosity $\nu = 0.005$ and $k_p = 40$. All runs are initially magnetically dominated with zero velocity field.}
    \label{tab:varying_Pm_v0.005_kp_40_zero}
\end{table*}

The plot in Figure \ref{fig:p_vs_Pm_v0.005} provides interesting clues about the behavior of the decay at different Pm. A common and perhaps obvious characteristic, is that both kinetic and magnetic decay become shallower as we increase the Prandtl number.  

For Pm $ \leq 0.25$, helical and nonhelical decays have the same decaying exponents, with the kinetic energy showing a shallower decay than the magnetic energy. In the range $2^{-5}\leq\text{Pm}\leq0.5$, $p_u$ remains approximately constant, whereas $p_b$ decreases. This is consistent with having a constant viscosity and decreasing resistivity. This indicates that the decay is mainly dominated by viscosity and diffusivity, with little influence of magnetic helicity and little interplay between the magnetic and kinetic fields. Nevertheless, it seems that there is a small transfer from magnetic to kinetic energy, that would explain why $p_b \gtrsim 5/2$ and $p_u \lesssim 5/2$, where $p=5/2$ is the expected value for a purely viscous or resistive decay \cite{batchelor1948decay,mansour1994decay}. 

For Pm $ > 0.5$ an interesting situation is observed. When we increase Pm, $p_b$ continues to decrease, and $p_u$, that was constant for Pm $<0.5$, starts decreasing. This indicates that the kinetic decay becomes shallower due to a more effective interaction with the magnetic field.

So far we have not made any distinction between the helical and nonhelical decay. In the nonhelical case, as we keep increasing Pm, we get $p_b < p_u$, thus, the magnetic decay becomes shallower than the kinetic one, with both exponents approaching 1. Nevertheless, the data shows no clear asymptotic behavior at this range of values. Studying the behavior at larger Pm would require a larger computational power that is not available at the moment. 

In the helical case, for $0.25 < $Pm$ \leq 1$ the magnetic decay is steeper than in the nonhelical case, but the kinetic decay is equal or slower, which suggest a more effective transfer of energy from magnetic to kinetic energy than in the nonhelical case. For Pm $ > 1$, both the kinetic and the magnetic energy decay at the same rate  $p_b\sim p_u$, and at a much slower rate than in the nonhelical case, approaching values in the range $\approx 1/2$ - $2/3$. The helical data seems to show the start of an asymptotic behavior at large Pm.

\begin{figure}
    \centering
    \includegraphics[width = \linewidth]{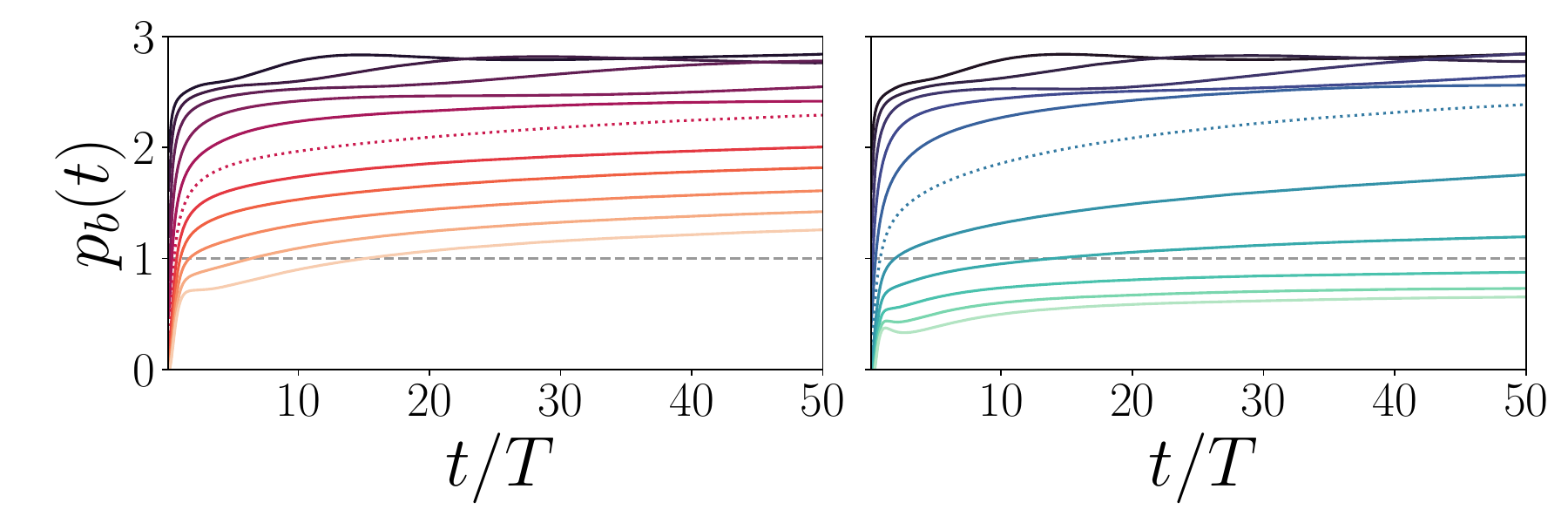}
    \caption{Time evolution of $p_b(t)$ for nonhelical (left) and helical (right) runs. Darker shades correspond to lower values of Pm, and dotted lines represent the case with Pm $ =1$.}
    \label{fig:p_vs_t_varying_Pm}
\end{figure}

 Now we look at the plots that show the evolution of these exponents in time, from which we take the measurements shown in the previous Figures. In Figure \ref{fig:p_vs_t_varying_Pm}, we see that in most cases, $p_b(t)$ seems to approach an asymptotic behavior. This indicates that there is a small bias towards smaller values in the measurements of $p_b$. For low Pm, $p_b(t)$ seems to approach an asymptotic value after a short transient, but it suffers a second transition at later times, where it ends up approaching a larger value. This is observed in both helical and nonhelical cases. This behavior is likely caused by the transition between turbulent and diffusive decay, which is reasonable at low Pm. The transients can be also influenced by the box size effect. 

\begin{figure}
    \centering
    \includegraphics[width = \linewidth]{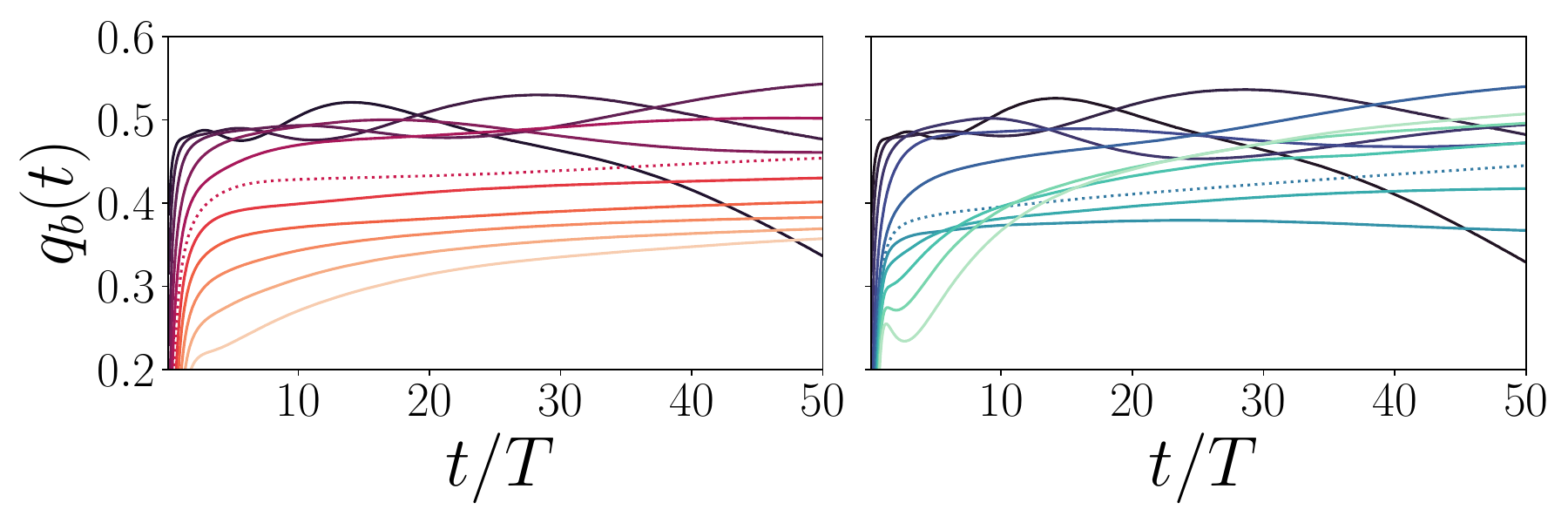}
    \caption{Time evolution of $q_b(t)$ for nonhelical (left) and helical (right) runs. Darker shades correspond to lower values of Pm, and dotted lines represent the case with Pm $ =1$.}
    \label{fig:q_vs_t_varying_Pm}
\end{figure}

We analyze the scaling laws of the coherent lengths $L_b$ and $L_u$. This exponent has an erratic behavior at low Prandtl number as we can see in Figure {\ref{fig:q_vs_t_varying_Pm}}. During an initial stage, $q_b(t)$ approaches an asymptotic behavior, but a few turnover times later, $q_b(t)$ starts to show an erratic behavior until it finally decays, without reaching an asymptotic behavior. At low Pm we expect a rather diffusive decay. With this initial spectrum, the decay is approximately of the form $E_b(k,t) = k^4 \exp\left(-2\eta k^2t\right)$, which gives $L_b \sim t^{1/2}$, however, the finite box size prevents $L_b$ from growing further, and this might explain the decay of $q(t)$ at later times.  

Because of the above reasons, the values of $q_b$ for Pm $<0.25$ should not be taken into account for the analysis. These are shown merely for the comparison with the behavior of $q_u$.

The measurements of $q_b$ are also taken at time $t\approx 50T$. The main characteristics shown in Figure \ref{fig:q_vs_Pm_v0.005}, is that $q_b$ and $q_u$ increase for increasing Pm in the helical case, whereas it decreases in the nonhelical case, showing that even though we observe a slight increase of magnetic energy at large scales in the nonhelical case, the coherence length does not grow significantly faster. The values of $q_b$ in the helical case seem to approach $1/2$ asymptotically for high Pm, whereas the nonhelical case does not show an asymptotic trend at this range.

\subsection{Scaling exponents in magnetically dominated decay}

\begin{figure*}[ht]
\centering
\subfloat[$p$ vs. Pm Mag. dominated]{
  \includegraphics[width=0.5\linewidth]{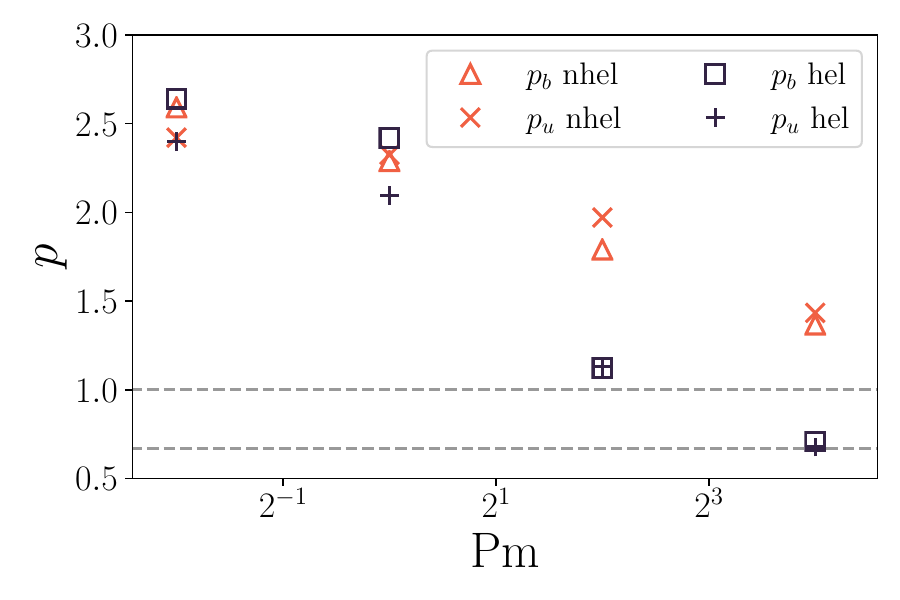}
  \label{fig:p_vs_Pm_v0.005_zero}
}
\subfloat[$q$ vs. Pm Mag. dominated]{
  \includegraphics[width=0.5\linewidth]{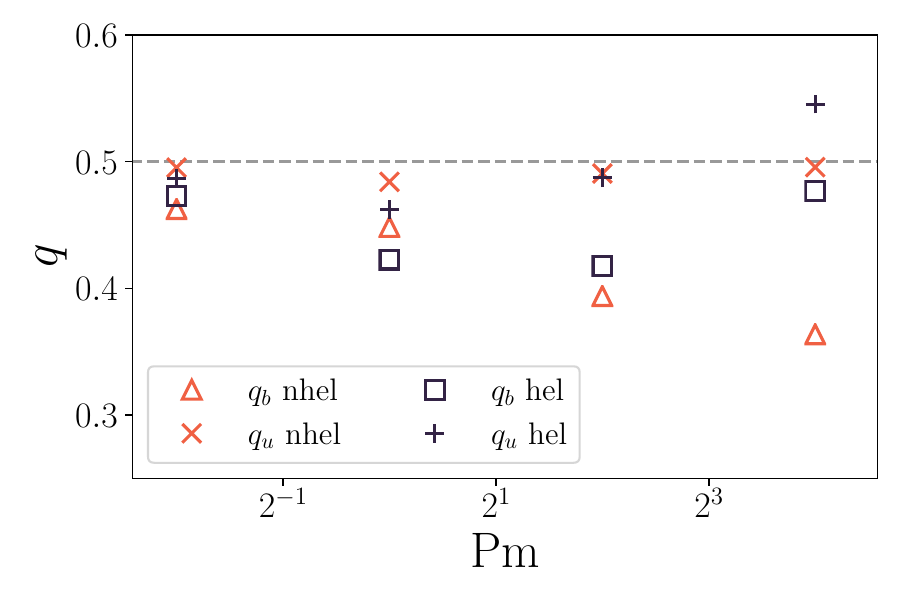}
  \label{fig:q_vs_Pm_v0.005_zero}
}
\caption{Scaling exponents $p$  (a) and $q$ (b) for the following cases: magnetic helical (blue squares), magnetic nonhelical (orange triangles), kinetic helical (blue crosses) and kinetic nonhelical (orange crosses). Dashed horizontal lines correspond to the typical scaling values observed in the literature $p=1$ for nonhelical flows and $q=2/3$ for helical flows. All runs are initially magnetically dominated.}
\label{fig:p_and_q_v0.005_zero}
\end{figure*}

We now look at the runs initialized with zero velocity. Figure \ref{fig:p_and_q_v0.005_zero} shows the dependence of the exponents on Pm in this case.

According to recent literature, magnetically dominated should give a stronger inverse cascade in the nonhelical case \cite{bhat2021inverse}. Nonetheless, our results show that the decaying exponents are surprisingly similar to the runs with initial equipartition. The only difference is that the values of $q_u$ in the nonhelical case are slightly larger than in equipartition cases. This indicates a faster growth of coherence length only in the kinetic field. The magnetic decay rate shows no relevant differences between magnetically dominated flows and those in equipartition, as opposed to what is stated in \cite{bhat2021inverse,hosking2021reconnection}.

\subsection{Spectra evolution}

\begin{figure*}[ht!]
\centering
\subfloat[NH\textsubscript{p-2} - Equipartition]{
  \includegraphics[width=0.25\linewidth]{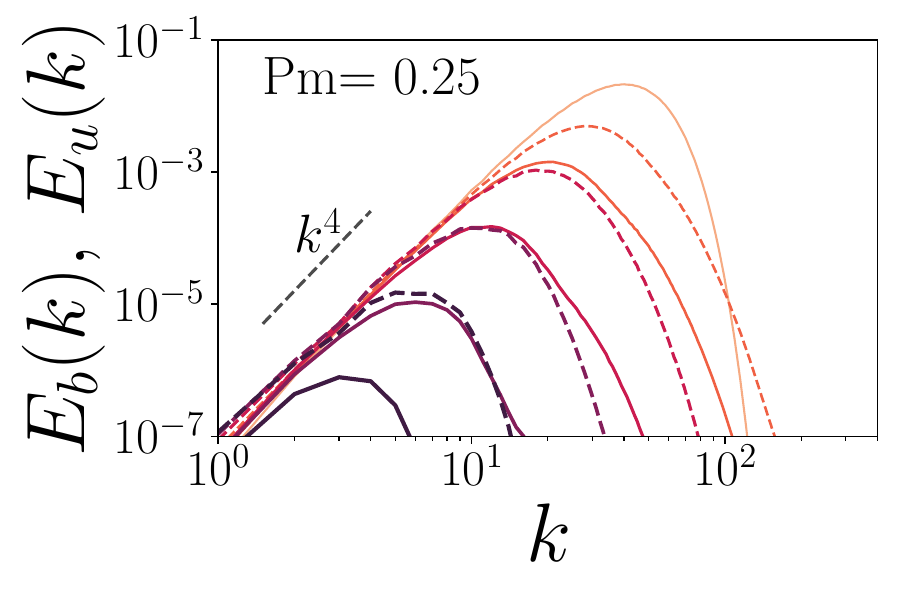}
  \label{fig:NHp-2}
}
\subfloat[NH\textsubscript{pz-2} - Mag. dominated]{
  \includegraphics[width=0.25\linewidth]{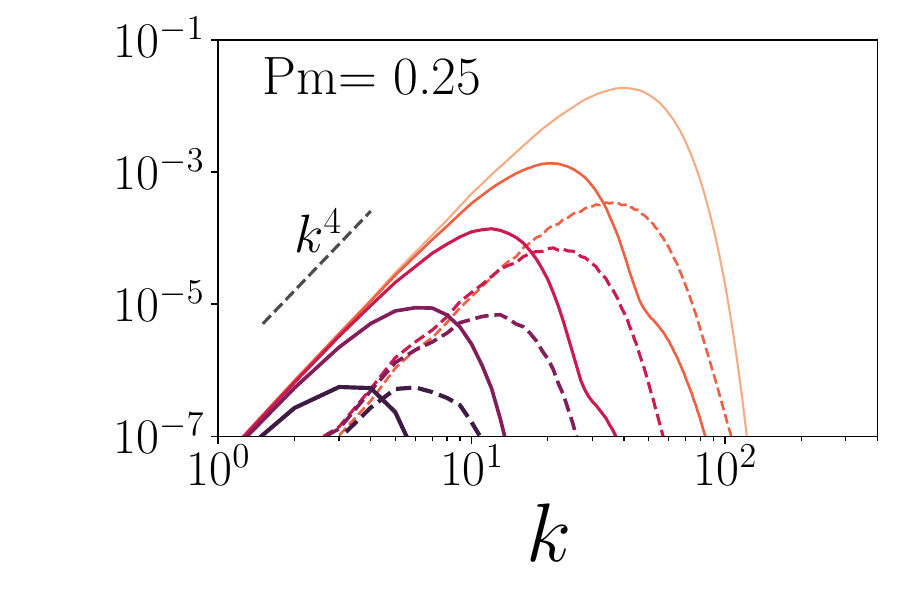}
  \label{fig:NHpz-2}
}
\subfloat[H\textsubscript{p-2} - Equipartition]{
  \includegraphics[width=0.25\linewidth]{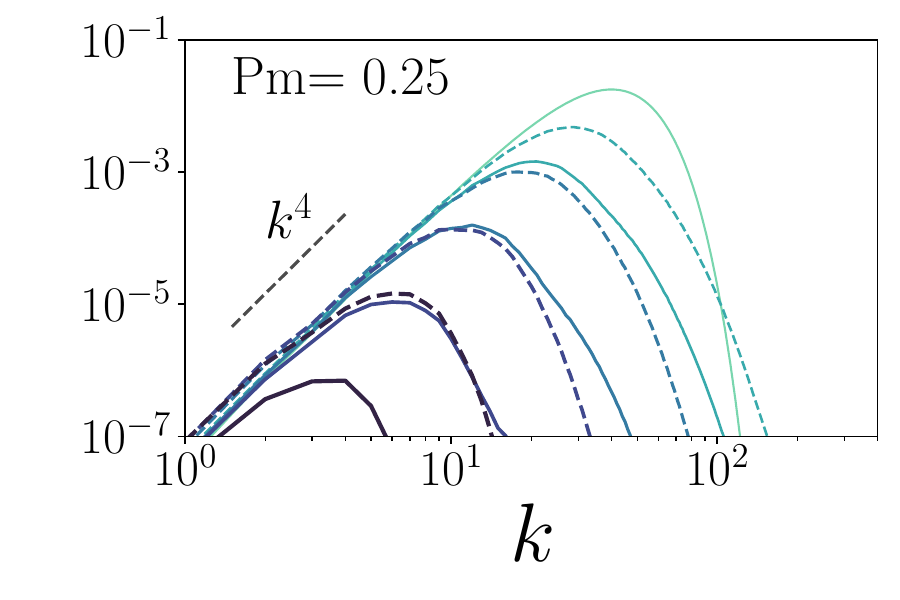}
  \label{fig:Hp-2}
}
\subfloat[H\textsubscript{pz-2} - Mag. dominated]{
  \includegraphics[width=0.25\linewidth]{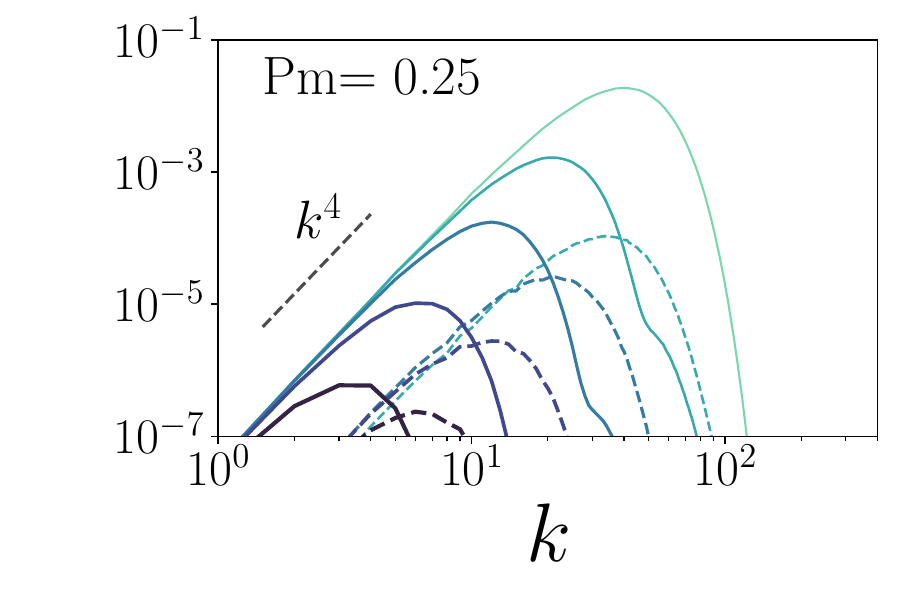}
  \label{fig:Hpz-2}
}
\hspace{0mm}
\subfloat[NH\textsubscript{p0} - Equipartition]{
  \includegraphics[width=0.25\linewidth]{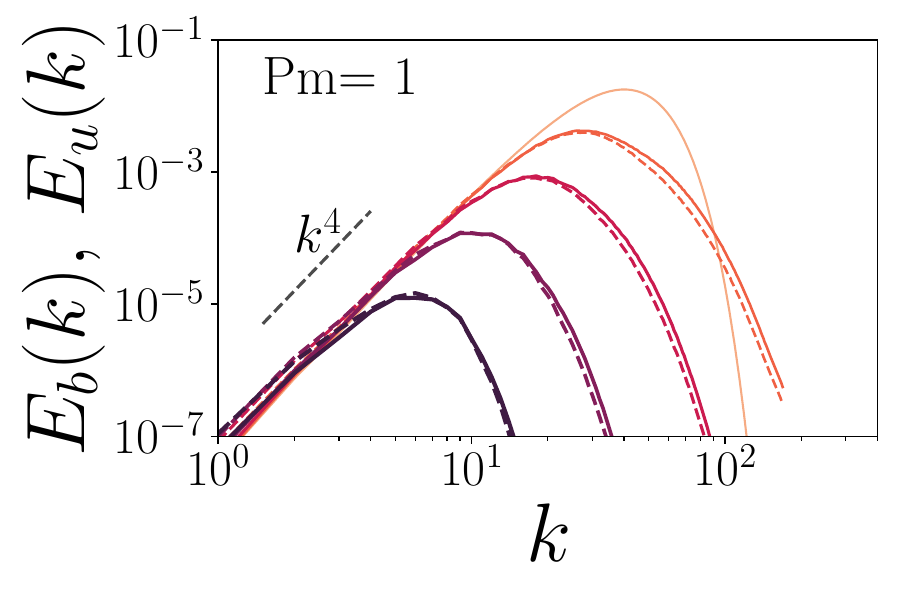}
  \label{fig:NHp0}
}
\subfloat[NH\textsubscript{pz0} - Mag. dominated]{
  \includegraphics[width=0.25\linewidth]{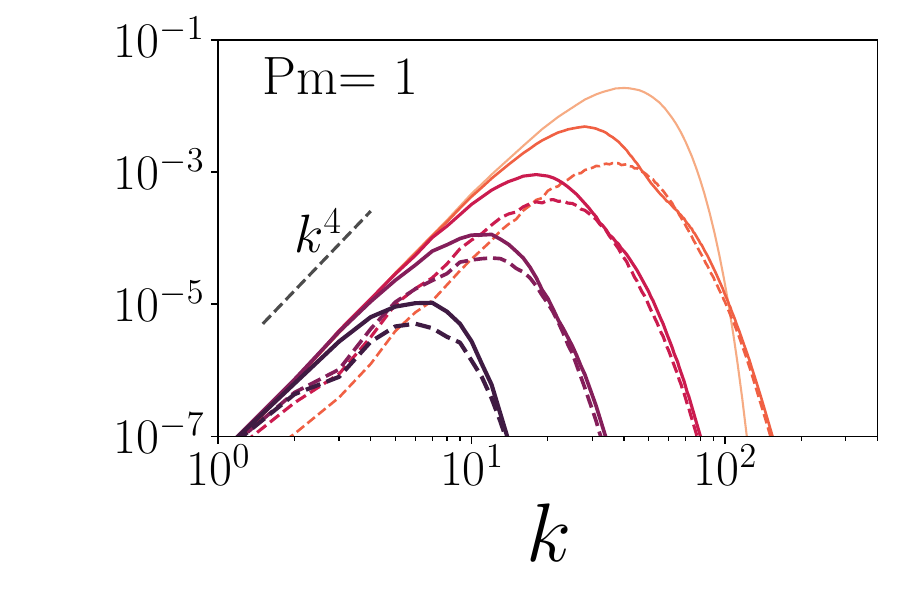}
  \label{fig:NHp0z}
}
\subfloat[H\textsubscript{p0} - Equipartition]{
  \includegraphics[width=0.25\linewidth]{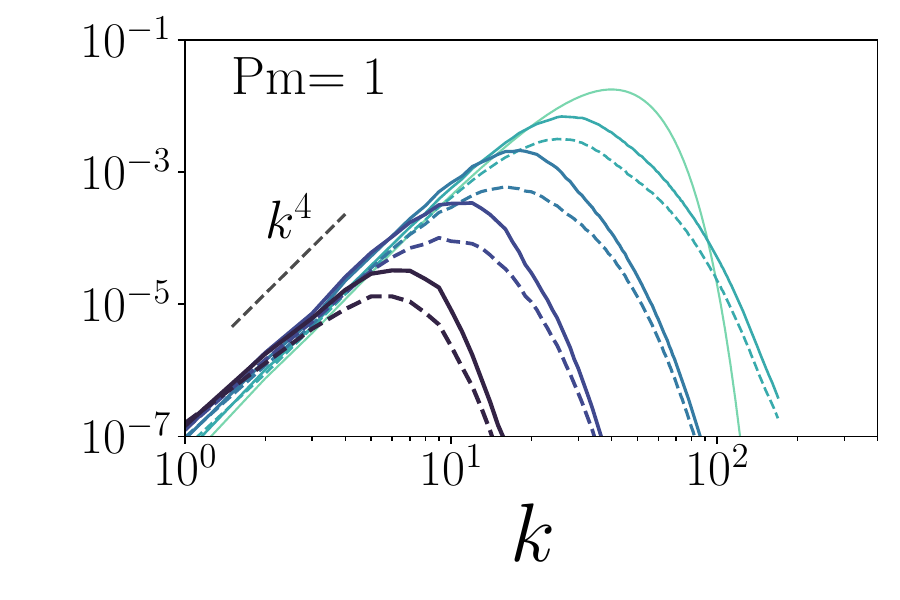}
  \label{fig:Hp0}
}
\subfloat[H\textsubscript{pz0} - Mag. dominated]{
  \includegraphics[width=0.25\linewidth]{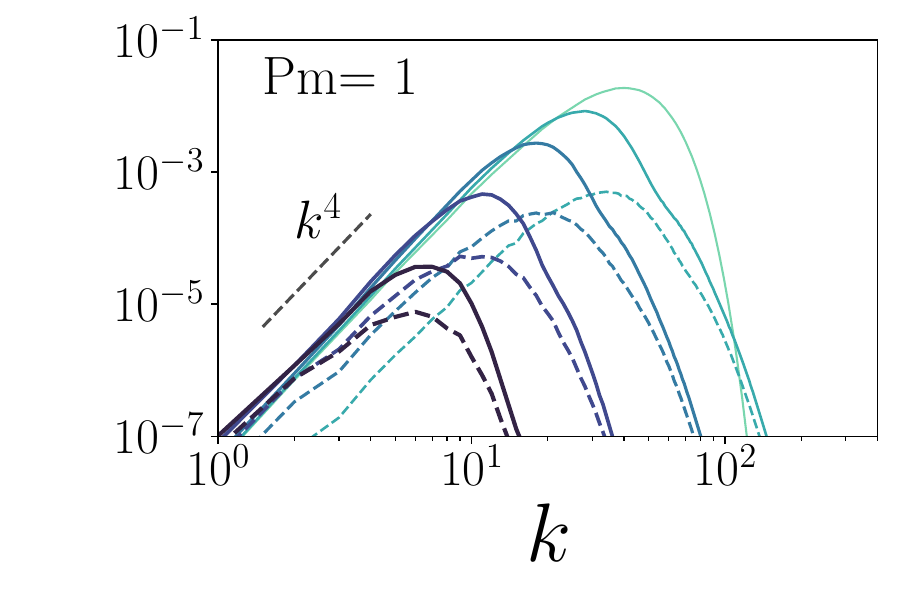}
  \label{fig:Hp0z}
}
\hspace{0mm}
\subfloat[NH\textsubscript{p2} - Equipartition]{   
  \includegraphics[width=0.25\linewidth]{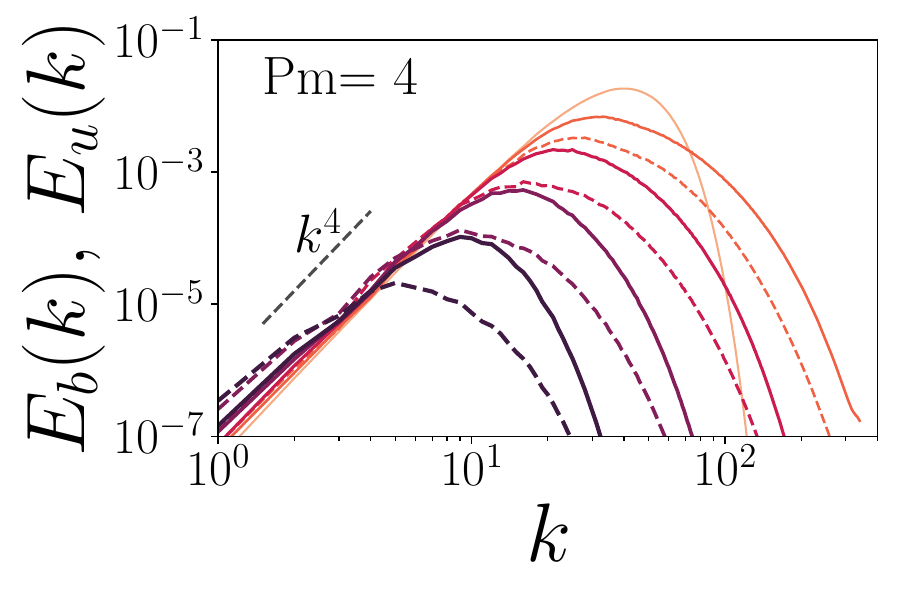}
  \label{fig:NHp3}
}
\subfloat[NH\textsubscript{pz2} - Mag. dominated]{   
  \includegraphics[width=0.25\linewidth]{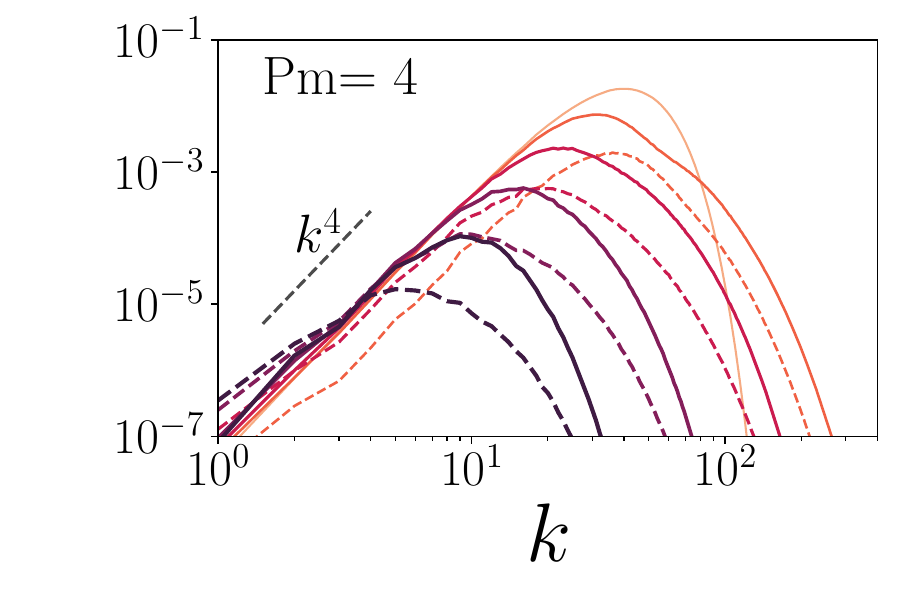}
  \label{fig:NHp3z}
}
\subfloat[H\textsubscript{p2} - Equipartition]{
  \includegraphics[width=0.25\linewidth]{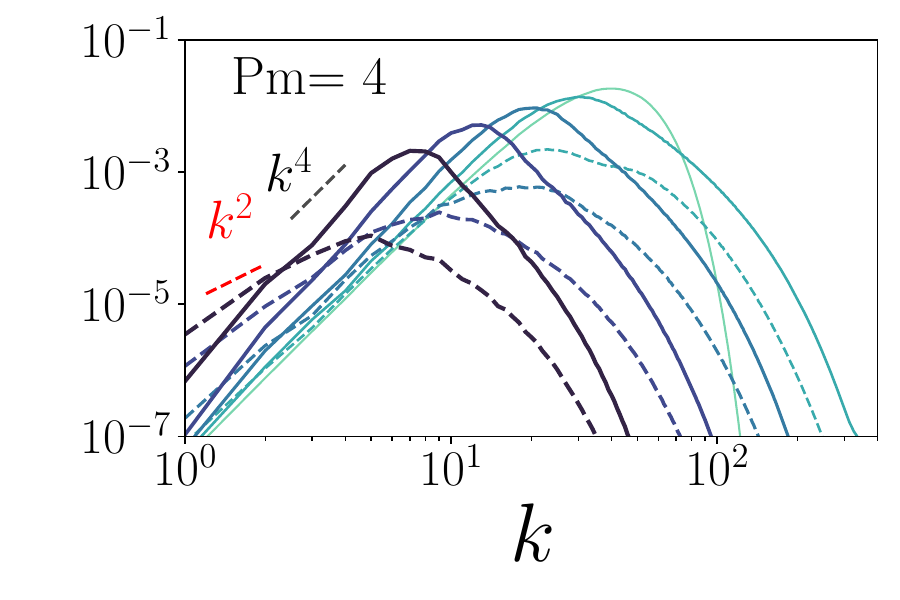}
  \label{fig:Hp3}
}
\subfloat[H\textsubscript{pz2} - Mag. dominated]{
  \includegraphics[width=0.25\linewidth]{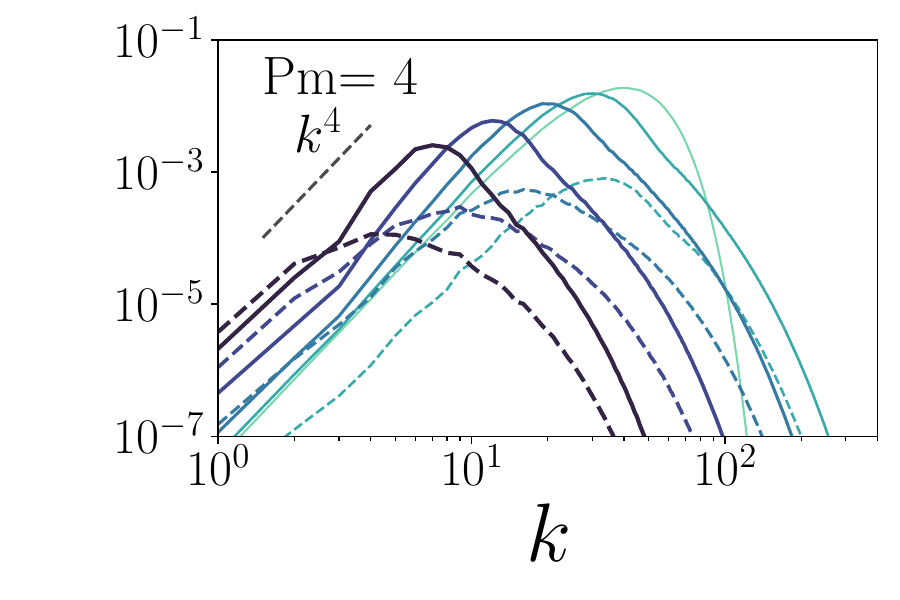}
  \label{fig:Hp3z}
}
\hspace{0mm}
\subfloat[NH\textsubscript{p4} - Equipartition]{
  \includegraphics[width=0.25\linewidth]{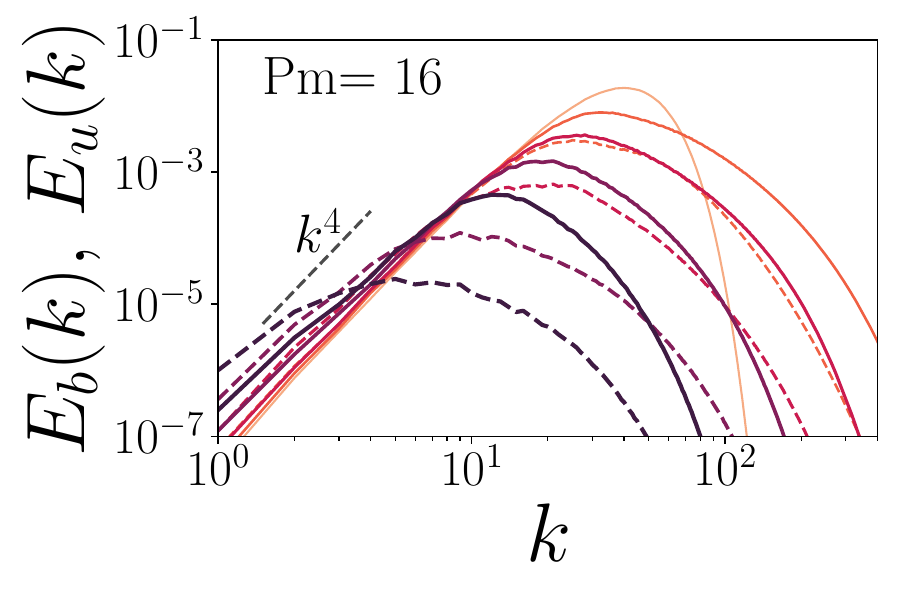}
  \label{fig:NHp5}
}
\subfloat[NH\textsubscript{pz4} - Mag. dominated]{
  \includegraphics[width=0.25\linewidth]{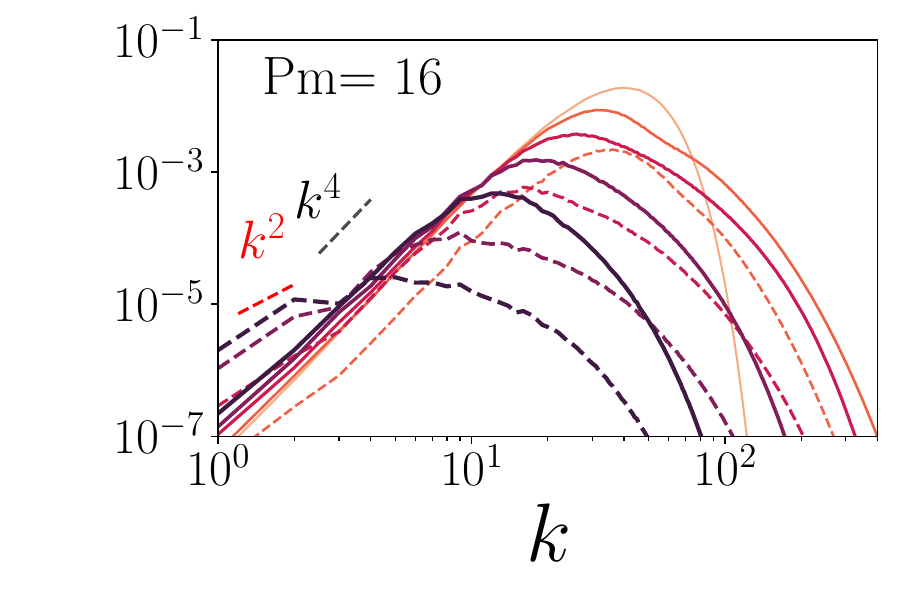}
  \label{fig:NHpz5}
}
\subfloat[H\textsubscript{p4} - Equipartition]{
  \includegraphics[width=0.25\linewidth]{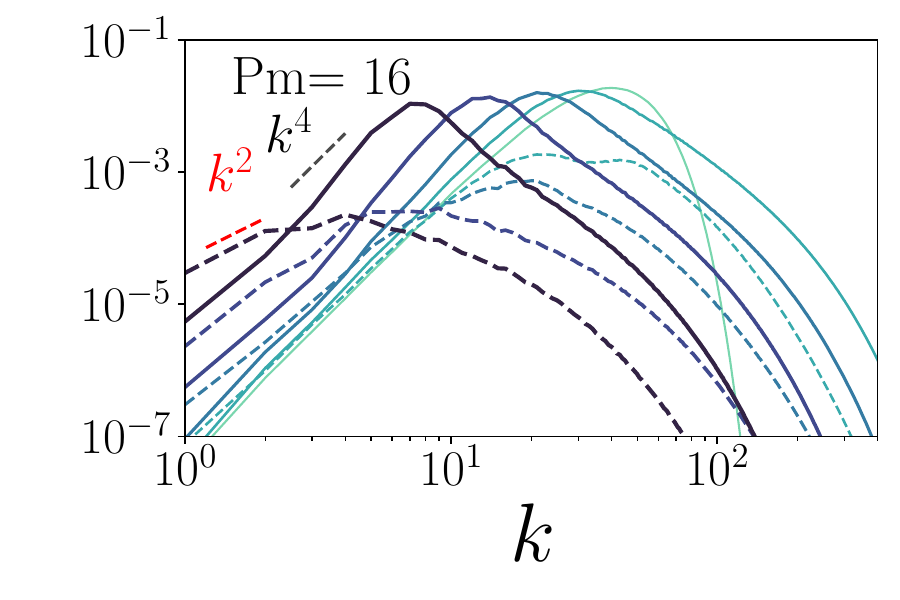}
  \label{fig:Hp5}
}
\subfloat[H\textsubscript{pz4} - Mag. dominated]{
  \includegraphics[width=0.25\linewidth]{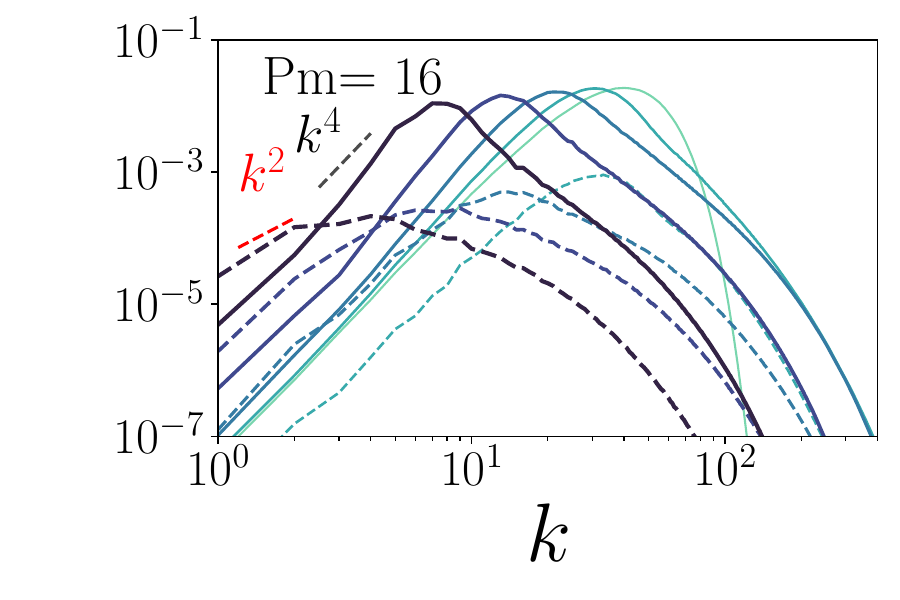}
  \label{fig:Hpz5}
}
\caption{Spectra evolution of runs NH/H\textsubscript{p-2,p0,p2,p4} and NH/H\textsubscript{pz-2,pz0,pz2,pz4}, for times $t/T = 0$, $0.8$, $3.3$, $13$ and $50$. Solid lines indicate magnetic energy spectra and dashed lines indicate kinetic energy spectra. Brighter lines correspond to earlier times.}
\label{fig:spectra_comparison_varying_Pm}
\end{figure*}

It is useful to compare all the previous cases by looking at the spectra evolution for varying Pm. 
Figure \ref{fig:spectra_comparison_varying_Pm} shows the spectra evolution for helical and nonhelical cases, and both magnetically dominated and in equipartition. Many of the features already mentioned are seen here with more clarity.

We see that, for Pm $ \ll 1$, the helical and nonhelical decays are almost identical. At Pm $ = 1$, we see that the magnetic decay is slower in the helical case. The kinetic spectra shows an identical decay in equipartition cases, and a more pronounced inverse transfer in the magnetically dominated case, due to the interaction with the magnetic field at large scales. 

For larger Pm, the inverse transfer of magnetic energy becomes pronounced in the helical case, as expected. The peak of the spectrum goes well past the initial $k^4$ spectrum. On the other hand, in the nonhelical case, only a weak build-up of magnetic energy is observed close to the characteristic wavenumber of the box, and the peak never goes past the initial $k^4$ spectrum, which indicates a slow growth of the coherence length. 

Interestingly, the kinetic energy shows a remarkable growth of energy at low wavenumbers, especially in the magnetically dominated nonhelical case and in both helical cases, adopting a $k^2$-like spectrum. The tilt of the large scale spectrum might be originated only due to the saturation at the box size. Nevertheless, this growth of kinetic energy indicates that even though viscosity is kept constant across simulations, the triadic interactions in the Lorentz force term make the inverse transfer of kinetic energy much more effective at high Pm. It is known that in the helical case, the inverse energy transfer is driven by the inverse cascade of magnetic helicity, which is better understood. In \cite{linkmann2016helical}, it is found that in the nonhelical case, transfers between different fields are more non-local than between the same fields. Furthermore, non-local triadic interactions contributing to the inverse transfer of energy are less constrained than those contributing to the forward transfer, especially in the magnetic dominated case, thus enhancing the inverse transfers. This might explain the larger inverse transfer of kinetic energy observed in the magnetic dominated case compared to the equipartition case in nonhelical flows.

\begin{figure}
    \centering
    \includegraphics[width=\linewidth]{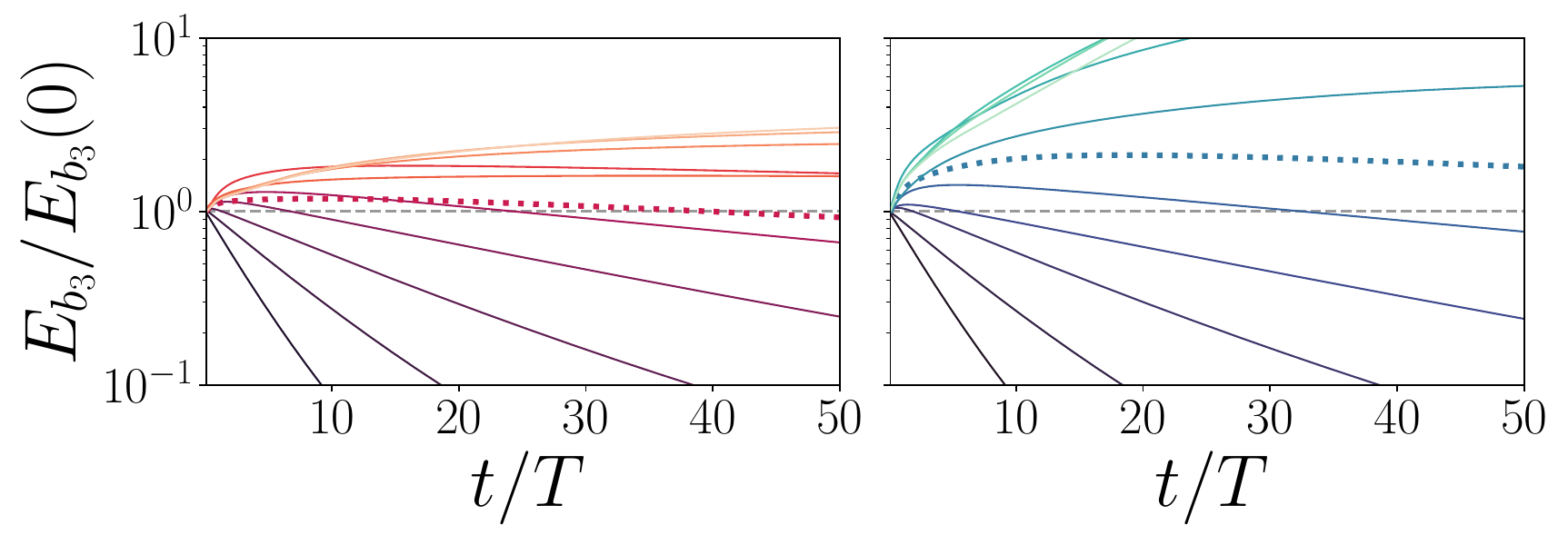}
    \caption{Large scale energy evolution $E_{b_3}(t)/E_{b_3}(0)$, for nonhelical (left) and helical (right) runs H/NH\textsubscript{p}. Brighter lines correspond to higher Pm. The dotted line in each plot corresponds to the run with Pm $=1$.}
    \label{fig:E3_vs_t}
\end{figure}

In Figure \ref{fig:E3_vs_t}, we show the build-up of magnetic energy at large scales in the equipartition case. We compute the energy at large scales (i.e.$E_{b_3}\int_0^3 dk\, E(k)$). We see that the build-up of large scale energy grows with Pm, although the initial transient becomes suddenly slower for the cases with Pm $ \geq 8$. The results are qualitatively similar in the magnetically dominated case. This is in line with the findings in \cite{mckay2019fully}. However, in \cite{reppin2017nonhelical} the opposite trend is found.

\subsection{High Reynolds number}

Last, we perform a similar analysis to the one done in section \ref{se:v_0.005_Pm}, but for a smaller viscosity. This allows us to explore the Pm variation at higher Re. The cost of this is that resolution limitations prevent us from exploring values of Pm larger than 1. Simulations parameters are shown in Table \ref{tab:varying_Pm_v0.0003125_kp_40}. All the cases analyzed in this section are initially in equipartition. 
\begin{table*}[t]
    \centering
   \begin{tabular}{llrrrr}
\toprule
\toprule
   Run &     Pm &  $\nu$ &  Re &  $k_p$ &   $N$ \\
\midrule
 H/NH\textsuperscript{*}\textsubscript{p-4} &  0.0625 &  0.0003125 &   128 &     40 &  2048 \\
 H/NH\textsuperscript{*}\textsubscript{p-3} &  0.125 &  0.0003125 &   128 &     40 &  2048 \\
 H/NH\textsuperscript{*}\textsubscript{p-2} &  0.25 &  0.0003125 &   128 &     40 &  2048 \\
 H/NH\textsuperscript{*}\textsubscript{p-1} &  0.5 &  0.0003125 &   128 &     40 &  2048 \\
 H/NH\textsuperscript{*}\textsubscript{p0} &  1&  0.0003125 &   128 &     40 &  2048 \\
\bottomrule
\bottomrule
\end{tabular}
    \caption{Helical and nonhelical runs for varying Prandtl number and fixed viscosity $\nu = 0.003125$ and $k_p = 40$.}
    \label{tab:varying_Pm_v0.0003125_kp_40} 
\end{table*}

\begin{figure*}[ht]
\centering
\subfloat[$p$ vs. Pm]{
  \includegraphics[width=0.5\linewidth]{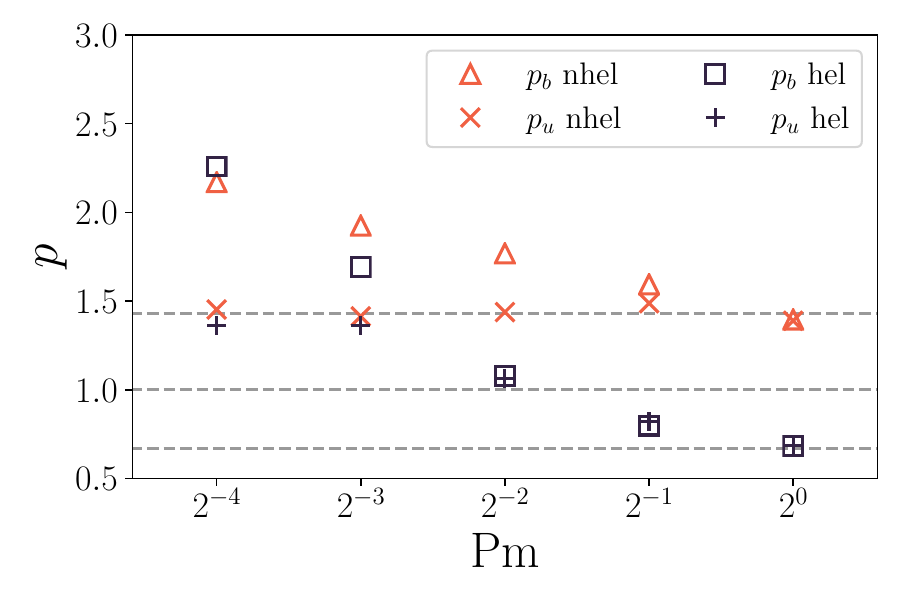}
  \label{fig:p_vs_Pm_v0.0003125}
}
\subfloat[$q$ vs. Pm]{
  \includegraphics[width=0.5\linewidth]{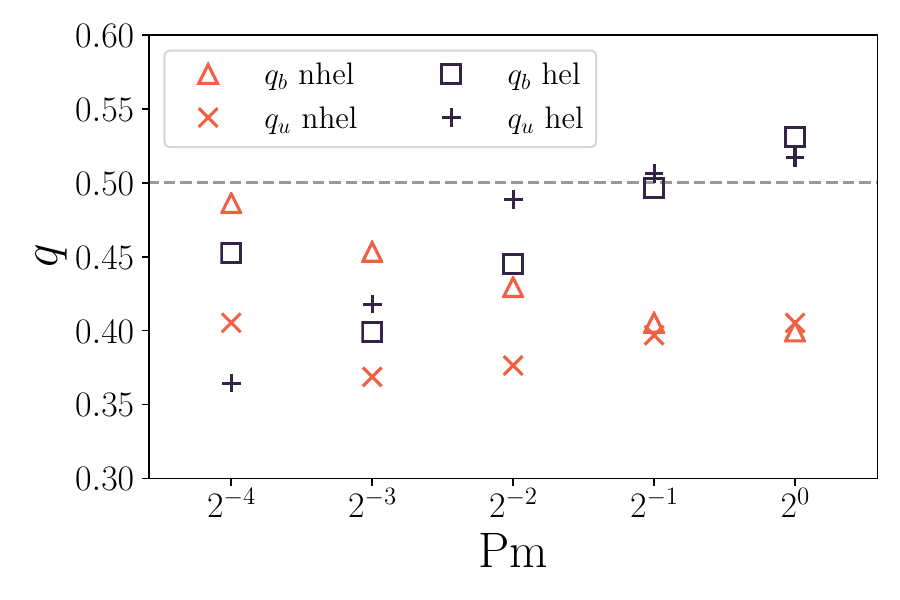}
  \label{fig:q_vs_Pm_v0.0003125}
}
\caption{Scaling exponents $p$  (a) and $q$ (b) for the following cases: magnetic helical (blue squares), magnetic nonhelical (orange triangles), kinetic helical (blue crosses) and kinetic nonhelical (orange crosses). Dashed horizontal lines correspond to the typical scaling values observed in the literature $p=1$ and $10/7$ for nonhelical flows and $q=2/3$ for helical flows. All runs are initially in equipartition.}
\label{fig:p_and_q_v0.0003125}
\end{figure*}

In Figure \ref{fig:p_and_q_v0.0003125}, we show the decaying exponent depenence on Pm. We compare this with the plot in Figure \ref{fig:p_vs_Pm_v0.005}, done for $\nu=0.005$. 

We see that some qualitative features remain similar, especially at low Pm. The magnetic decay is mainly diffusive and steeper than the kinetic. However, in this case the kinetic decay is turbulent, with a rate $p_u \approx 10/7$, which is reasonable given the higher Re in these simulations.

We note that the decay in the helical case is shallower than the nonhelical for Pm $ \geq 2^{-4}$. This is different to the case $\nu=0.005$, where the same occurs at Pm $ \geq 1$. Nevertheless, we note that in both cases, this corresponds to the the same magnetic Reynolds number Re$_b= $Re, Pm $\approx 8$, indicating that the helical inverse transfer becomes more effective as soon as the magnetic field starts to develop turbulence, independently of the kinetic field. 

The nonhelical kinetic decay does not become shallower as we increase Pm, indicating that the interaction with the magnetic field does not alter the evolution of the kinetic decay at this range of Pm. The magnetic and the kinetic exponents approach a Loitsyanksy-Kolmogorov decay of $p=10/7$. Interestingly, for equal values of resistivity, the magnetic decay is shallower in the case with lower Re.

In the helical case, the kinetic decay is strongly influenced by the interaction with the magnetic field as we increase Pm. Both end up decaying at the same rate for the largest values of Pm, similar to the high viscosity case. This shows that helical magnetic fields are more effective at sustaining the kinetic field than nonhelical ones.

Now we focus on the behavior of $q_b$. The trends are more defined at lower Pm than in Figure \ref{fig:q_vs_Pm_v0.005}, indicating that the erratic behavior observed is a consequence of low Reynolds numbers. For the largest values of Pm, the kinetic and magnetic integral lengthscales grow at the same rate, with the helical showing a value of $q_b \gtrsim 0.5$ and the nonhelical $q \approx 0.4$.

Even though the inverse transfer we observe is small, the fact that it increases with Pm is appealing in the context of possible cosmological applications, given the high values of Pm in that context. Unfortunately, running fully resolved DNS at higher Pm and Re is unfeasible at present. 

\subsection{Helicity conservation}
\label{se:Helicity conservation}
\begin{figure*}[ht]
    \centering
    \includegraphics[width=\textwidth]{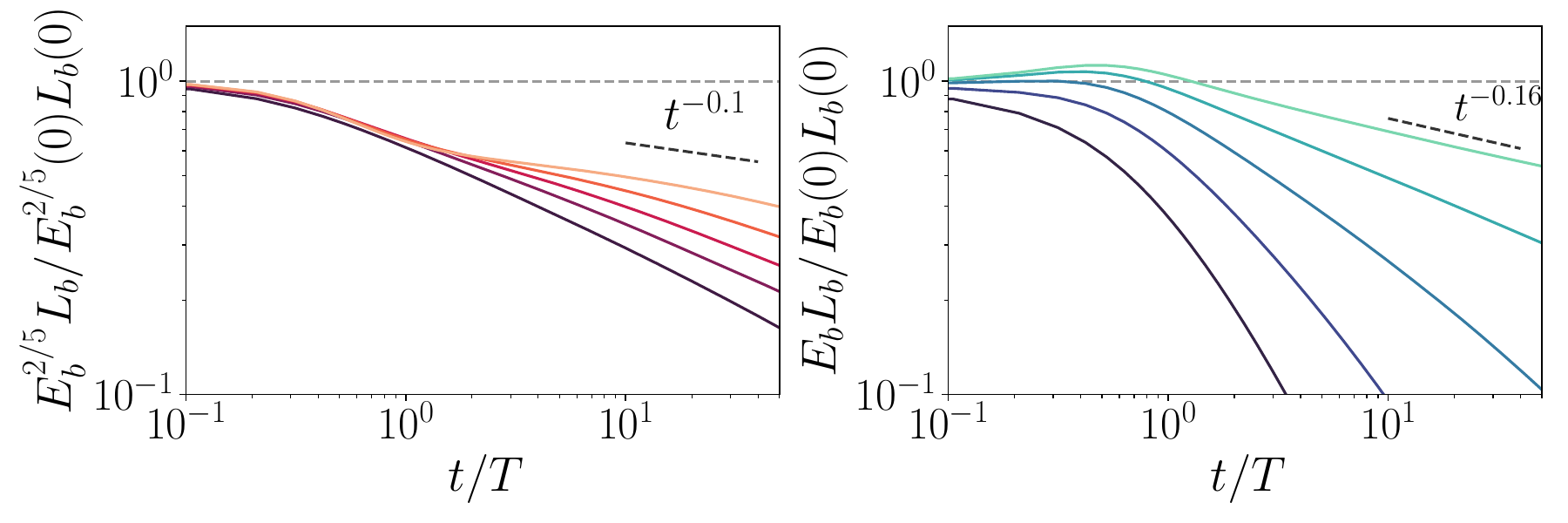}
    \caption{Time evolution of $E_b^{\beta} L_b/(E_b^{\beta}(0)L_b(0))$ for runs in Table \ref{tab:varying_Pm_v0.0003125_kp_40}. Where $\beta=1$ for helical runs and $\beta = 2/5$ for nonhelical runs.}
    \label{fig:helicity_and_Ih_conserv}
\end{figure*}

It is expected that at sufficiently large Reynolds and Lundquist numbers, magnetic helicity remains approximately conserved during the decay. In section \ref{se:decaying_turbulence}, we saw that this leads to a conservation of the form $E_bL\sim \text{const.}$ (see Eq. ( \ref{eq:Helicity_conserv2})). We also mentioned the analogous conservation predicted recently for nonhelical flows, which is of the form $E_b^2L^5$. In Figure \ref{fig:helicity_and_Ih_conserv}, we show the evolution of these quantities in time for runs in Table \ref{tab:varying_Pm_v0.0003125_kp_40}. We note that the helical case shows an approximate conservation in both cases, with $E_b L \sim t^{-0.16}$ and $E_b^{2/5} L \sim t^{-0.10}$ for Pm$=1$. 

The conservation is not exact, but there is a slow decay over time. We can assume that the decay follows a power law $E_b^{\beta}L \sim t^{-\delta}$. For $\delta =0$, it is possible use a Kolmogorov-like argument and show that, in the Sweet-Parker regime, the energy decay is given by $\partial_t E_b \sim E_b^{(6\beta + 5)/4}$ \cite{hosking2021reconnection}. This regime is typical in most DNS simulations of MHD turbulence, as the Lundquist number is high but not enough to reach the fast-reconnection regime \cite{uzdensky2010fast}. The Sweet-Parker regime is characterized by a balance between the resistive and inductive terms in Eq. (\ref{eq:MHDeqnsb}) \cite{parker1957sweet,sweet195814}. This gives a slower timescale for the reconnection than in the ideal case where the inductive term dominates. For $\delta \neq 0$, we get 

\begin{align}
    \partial_t E_b \sim E_b^{(6\beta + 5)/4} t^{-\delta}\quad, \\
    E_b \sim t^{-4(\delta+1)/(6\beta+1)}
\end{align}

This would produce a slight change in the predicted values of $p_b = (-6\beta-1)/4)$, such that the corrected values are now $p_b' = p_b(1+\delta)$. For a value of $\delta \approx 0.1$, the decay is slightly steeper by a factor of $10\%$. This is slight but significant when it comes to compare different theoretical predictions. In Fig. \ref{fig:p_vs_Pm_v0.0003125}, we saw that $p_b$ was close to $10/7$, but if we take this correction into account, the actual value would indicate that $p_b$ gets closer to the predicted values of $1.18$ or $1$. Once again, we remark the importance of these kind of details when comparing theoretical predictions. 

\section{Varying Reynolds number}
\label{se:varying_Re}

We study the behavior of the decay at varying Re with fixed Pm $=1$. This analysis is similar to the one done in \cite{berera2014magnetic}, in which the the same eddyBurgh code and same initial conditions were implemented. The only difference is the range of Reynolds numbers explored and the scale separation. In \cite{berera2014magnetic}, the authors find $p_b \approx 0.47 + 13.9 R_{\lambda}$, where $R_{\lambda}$ refers to the Taylor scale Reynolds number. 

We run a set of 5 helical and 5 nonhelical simulations to observe the behavior of the decay at varying Re with the parameters shown in Table \ref{tab:varying_Re_Pm1_kp_40}.

\begin{table*}[ht]
    \centering
    \begin{tabular}{lrrrrr}
    \toprule
    \toprule
    Run &  Pm &  $\nu$ &  Re &  $k_p$ &   $N$ \\
    \midrule
    H/NH\textsubscript{Re5} &      1 &  0.0003125 &   129 &     40 &  2048 \\
    H/NH\textsubscript{Re4} &      1 &  0.000625 &    64.5 &     40 &  2048 \\
    H/NH\textsubscript{Re3} &      1 &  0.00125 &    32.4 &     40 &  1024 \\
    H/NH\textsubscript{Re2} &      1 &  0.0025 &    16.2 &     40 &  1024 \\
    H/NH\textsubscript{Re1} &      1 &  0.005 &     8.09 &     40 &  1024 \\
    \bottomrule
    \bottomrule
\end{tabular}
    \caption{Helical and nonhelical runs for varying Re and fixed Prandtl number Pm $ = 1$ and $k_p = 40$. Darker shades correspond to higher Re. All runs are initially in equipartition.}
    \label{tab:varying_Re_Pm1_kp_40}
\end{table*}

\begin{figure}
    \centering
    \includegraphics[width=\linewidth]{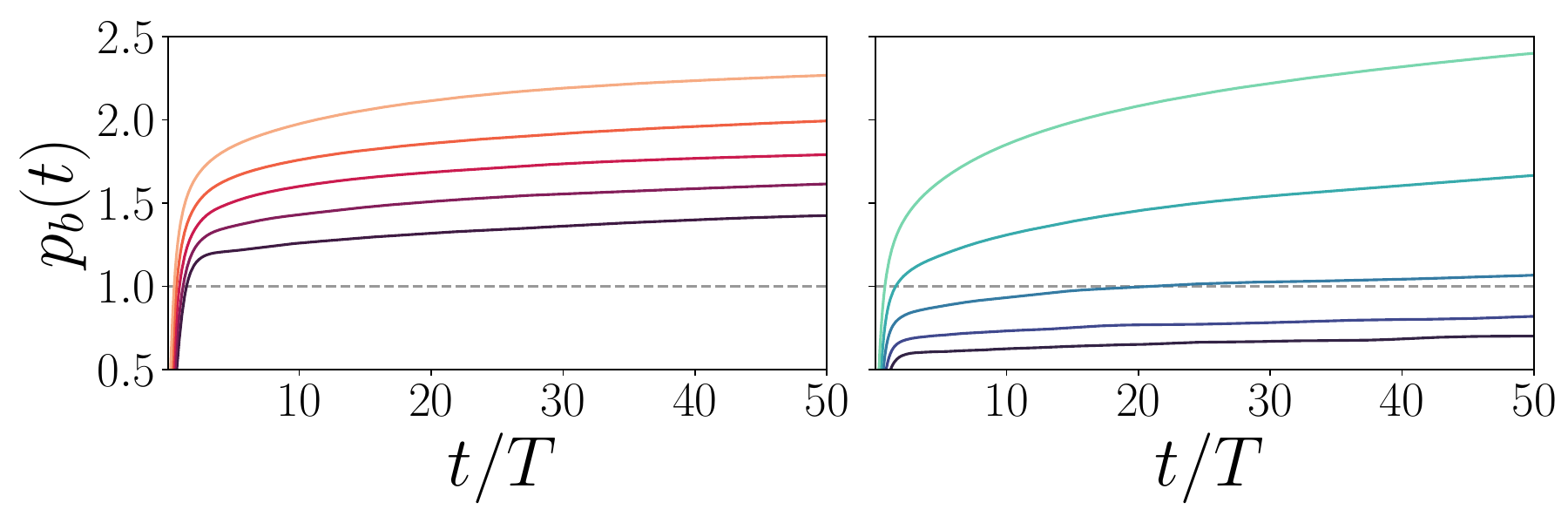}
    \caption{Evolution of $p_b(t)$ for nonhelical (left) and helical (right) flows at Pm $=1$ and varying Re. Darker shades correspond to higher Re. }
    \label{fig:p_vs_t_Pm1_varying_Re}
\end{figure}

The plot in Figure \ref{fig:p_vs_t_Pm1_varying_Re} shows the evolution of $p_b(t)$. All curves show a small but consistent growth that seem to approach an asymptotic value, except from the helical ones with lower Re in which the growth is still considerable. The measurements are taken using a linear fit in a log-log scale between $t/T=40$-$50$.

\begin{figure*}[ht]
\centering
\subfloat[$p$ vs. Re]{
  \includegraphics[width=0.5\linewidth]{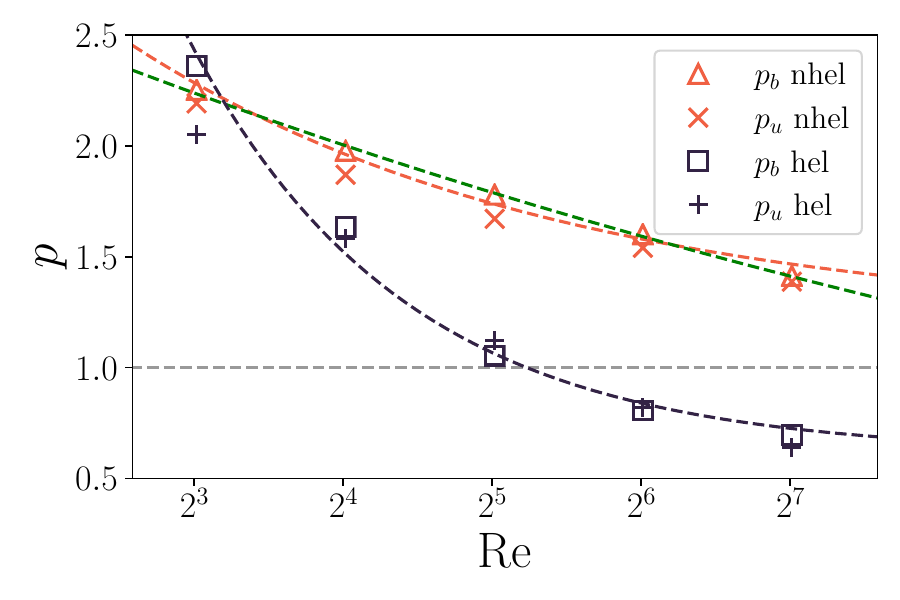}
  \label{fig:p_vs_Re_Pm1}
}
\subfloat[$q$ vs. Re]{
  \includegraphics[width=0.5\linewidth]{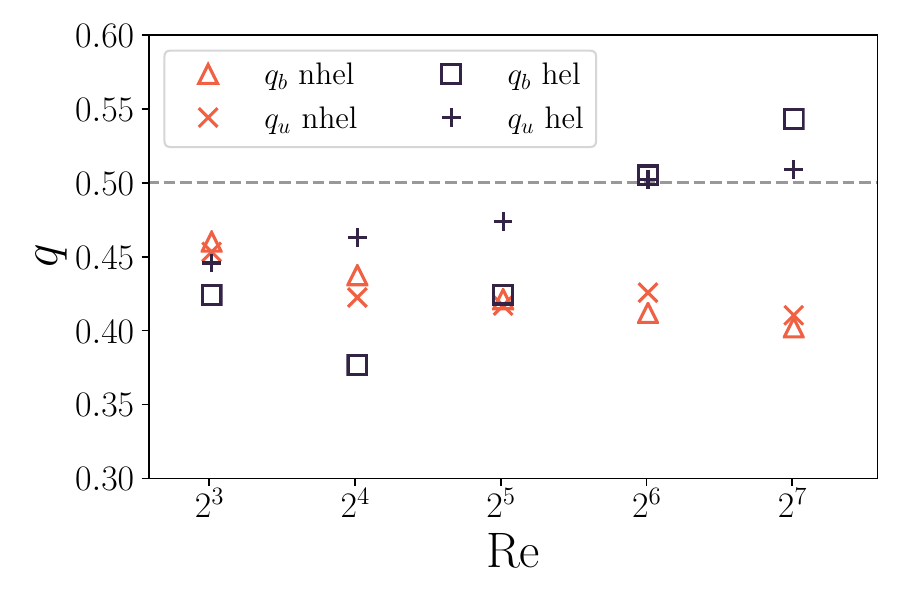}
  \label{fig:q_vs_Re_Pm1}
}
\caption{Scaling exponents $p$  (a) and $q$ (b) for the following cases: magnetic helical (blue squares), magnetic nonhelical (orange triangles), kinetic helical (blue crosses) and kinetic nonhelical (orange crosses). Dashed colored curves correspond to the fits $p_h(\text{Re})$ (blue) and $p_{nh}(\text{Re})$ (orange for $\alpha = 1/2$ and green for $\alpha=1/8$), and dashed horizontal lines correspond to some typical scaling values observed in literature $p=1$ for nonhelical flows and $q=1/2$ for helical flows. All runs are initially in equipartition.}
\label{fig:p_and_q_Pm1}
\end{figure*}

The measured values of $p$ and $q$ are shown in Figure \ref{fig:p_and_q_Pm1}, where only the helical $p_b$ and $p_u$ seem to approach an asymptotic value for large Re. We find that these have a dependence of the form $p_{b}($Re$) = p_{h\infty} + p_{h0}/\text{Re}$, with measured values of  $p_{h\infty} = 0.6 $ and $p_{h0} = 14 $. The asymptotic value is different to the one measured in \cite{berera2014magnetic}, possibly due to the wider range of Re explored in this work. This is essential to determine the asymptotic behavior. 
We can see that the value of $p_b$ is close to different predictions $p_b \approx 0.5-0.66$, and a similar behavior is observed for the kinetic decay, which is not in agreement with the prediction that $p_u\approx 1$ for helical flows in equipartition \cite{hosking2021reconnection}.

For the nonhelical case, in principle one can propose a similar dependence of the form $p_{b}($Re$) = p_{nh\infty} + p_{nh0}/\text{Re}^{\alpha}$. Nevertheless, the range of Re explored does not show a clear asymptotic behavior, and as a consequence, different fits with different values of $\alpha$ fit the data, giving quite different values of $p_{nh\infty}$. For instance, $p_{nh\infty} = 1.2$ for $\alpha=1/2$ and $p_{nh\infty} = -0.6$ for $\alpha =1/8$, which is unphysical. Both fits are shown in the plot of Figure \ref{fig:p_vs_Re_Pm1}.

Even obtaining a clear asymptotic behavior, these results depend strongly on the measurement methods. The lack of accuracy makes it difficult to distinguish between different theoretical predictions. For instance, between the $E_b \sim t^{-1.11}$ and $t^{-1.18}$ scalings given by the fast and slow reconnection regime respectively. Also, an ensemble average using different initial conditions would give a better estimation of the error.

\begin{figure*}
    \centering
\subfloat[Equipartition]{
    \includegraphics[width=0.5\linewidth]{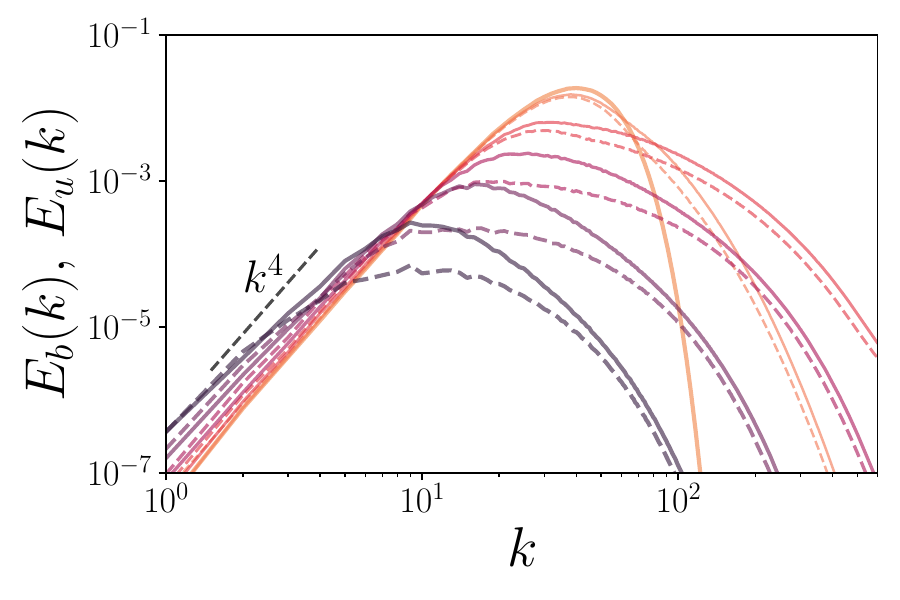}
    \label{fig:Eb_nh_equi}
}
\subfloat[Mag. dominated]{
    \includegraphics[width=0.5\linewidth]{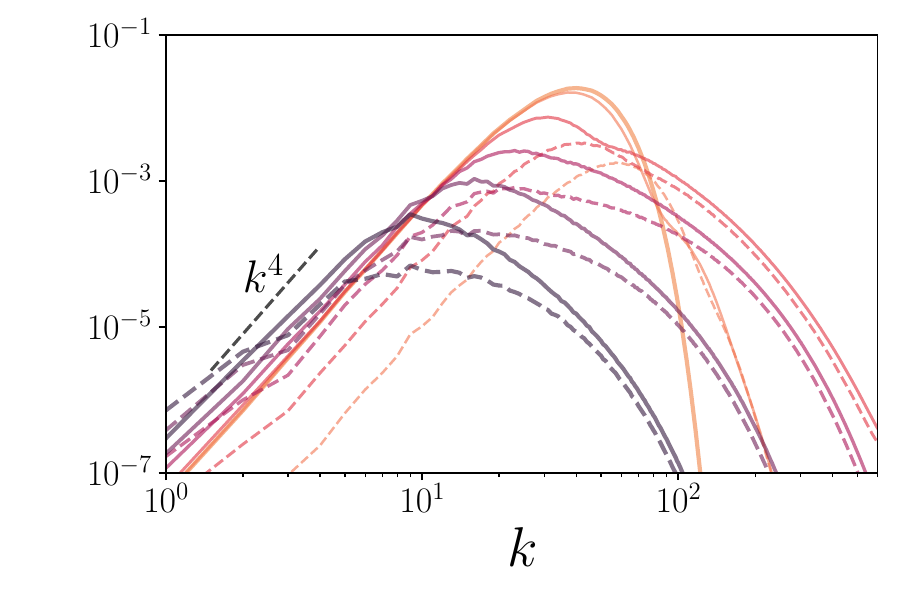}
    \label{fig:eb_nh_zero}
}
     \caption{Spectra evolution of runs NH\textsubscript{Re5}, and NH\textsubscript{Re5Z}, for times $t/T = 0$,$0.2$, $0,8$, $3$, $13$, and $52$. Solid curves represent the magnetic energy spectra and dashed curves represent the kinetic spectra.}
     \label{fig:spectra_equi_vs_zero}
\end{figure*}

Last, we consider the run NH\textsubscript{Re5}, that has the largest Re, and we run another with the same initial magnetic field but with the velocity field initialized to zero. Figure \ref{fig:spectra_equi_vs_zero} shows the spectra evolution in both cases.

We see that for the magnetically dominated case, the kinetic spectra has a pronounced inverse transfer and ends up with more energy at large scales than the run that starts in equipartition. 

\begin{figure}
    \centering
    \includegraphics[width=\linewidth]{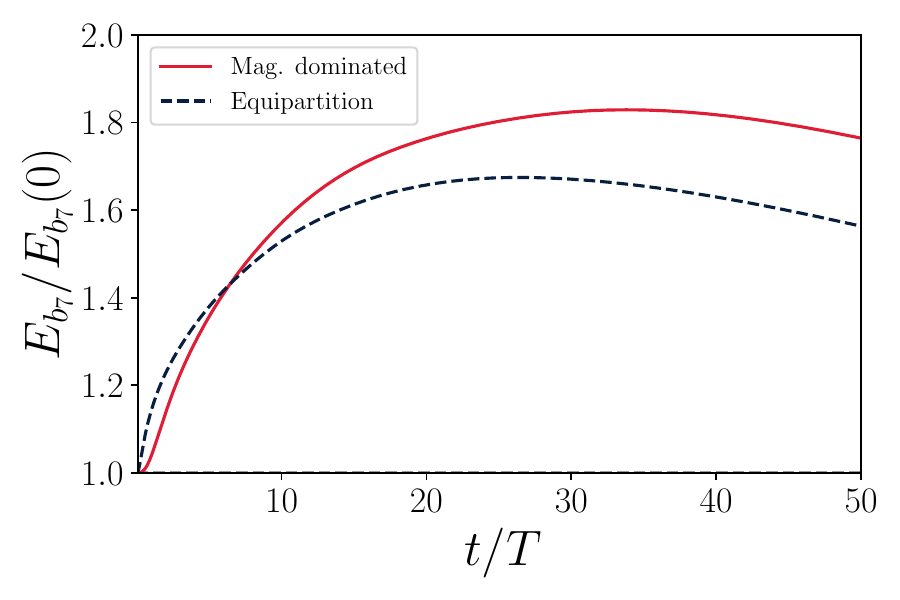}
    \caption{Time evolution of the large scale magnetic energy $E_{b_3}(t)/E_{b_3}(0)$ for runs NH\textsubscript{Re5}, and NH\textsubscript{Re5Z}.}
    \label{fig:Eb7}
\end{figure}

The comparison of the magnetic spectra is less clear. For that reason, we look at the large scale energy up to $k=7$. This is shown in Figure \ref{fig:Eb7}. The magnetically dominated case shows a slightly larger growth of magnetic energy at large scales. This is, at least qualitatively, in line with the analysis in \cite{bhat2021inverse}. 

\section{Varying $k_p$}
\label{se:varying_kp}

Now we perform a study at varying $k_p$. We can think of a set of simulations in which we keep the initial energy, viscosity and resistivity constant, but we vary the value of $k_p$. If we choose a value of $k_p \gg 1$, we obtain enough scale separation to allow inverse energy transfer without being affected by the box size, the setback is that by doing this, most of our energy will be close to the dissipative scales, hence, most of the decay will be dissipative and no inertial range will develop. If we choose $k_p \gtrsim 1$, we will get a wider inertial range, but without enough scale separation to allow inverse transfer and avoid box size effects. For this section we concentrate only in the nonhelical case. 

To establish a fair comparison between runs, we perform a set of 6 simulations with $k_p$ ranging from 5 to 80, and choosing $\nu$ such that the Reynolds number is constant for all runs. Simulations parameters are shown in Table \ref{tab:varying_kp_fixedRe}.

\begin{table*}[ht]
    \centering
\begin{tabular}{lrrrrr}
\toprule
\toprule
                     Run &  Pm &    $\nu$ &  Re &  $k_p$ &   $N$ \\
\midrule
NH\textsubscript{$k_p5$}  &      1 &  0.01 &  32.2 &      5 &   256 \\
NH\textsubscript{$k_p10$} &      1 &  0.005 &  32.2 &     10 &   512 \\
NH\textsubscript{$k_p20$} &      1 &  0.0025 &  32.2 &     20 &   512 \\
NH\textsubscript{$k_p40$} &      1 &  0.00125 &  32.2 &     40 &  1024 \\
NH\textsubscript{$k_p80$} &      1 &  0.000625 &  32.2 &     80 &  2048 \\
\bottomrule
\bottomrule
\end{tabular}
    \caption{Nonhelical runs varying $k_p$ at fixed Reynolds number Re$(0) \approx 32$.}
    \label{tab:varying_kp_fixedRe}
\end{table*}

\begin{figure}
    \centering
    \includegraphics[width=\linewidth]{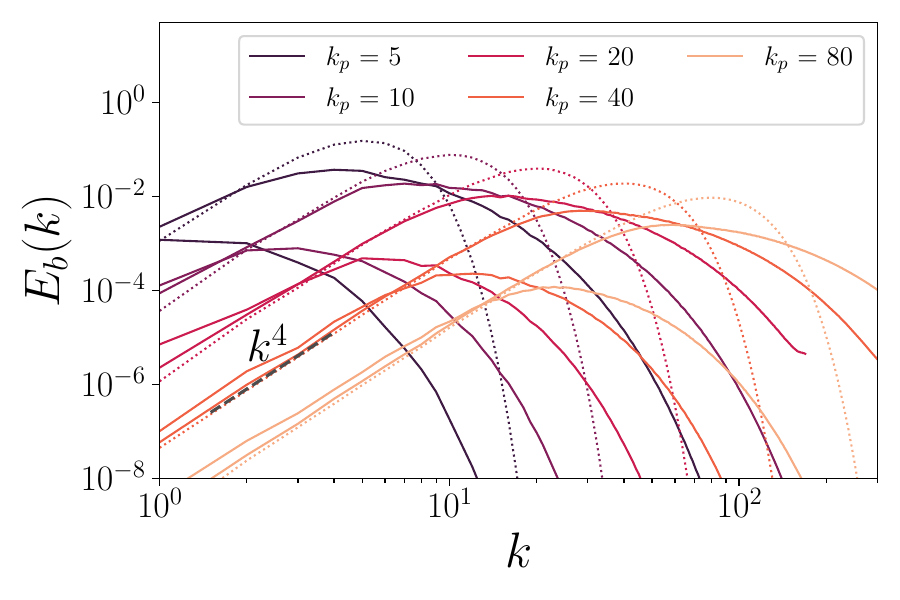}
    \caption{Spectra evolution for times $t/T = 0, 1 $ and $20$ for $k_p=20,40$ and $80$. Keeping Pm $=1$ and Re$ \approx 32$. }
    \label{fig:fixed_Re_spectra}
\end{figure}

We start by taking a look at the spectra evolution in Figure \ref{fig:fixed_Re_spectra}. We focus on the behavior for wavenumbers $k < k_p$. We note that the small growth of energy in the low $k$ region keeps the initial $k^4$ form in cases with $k_p \geq 40$. Instead, for $k_p <40$, box size effects start to become noticeable, and saturation tilts the $k^4$ spectra towards shallower slopes. Still, it is worth noticing that in every case, the decay produces a small growth of energy at large scales. 

\begin{figure}
    \centering
    \includegraphics[width=\linewidth]{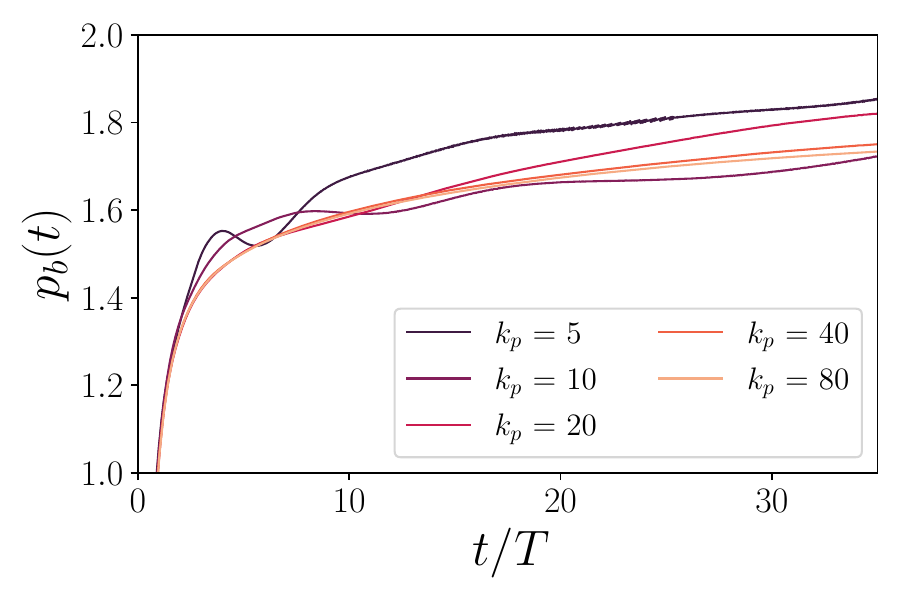}
    \caption{Evolution of $p_b(t)$ for nonhelical runs with fixed Re$ \approx 32$, Pm $=1$, and varying $k_p = 5,10,20,40,80,$ and $160$. }
    \label{fig:p_vs_t_fixedRe}
\end{figure}

We measured the decaying exponents and we find no drastic differences for all values of $k_p$, with $p_b\approx 1.7 $-$1.9$. Nevertheless, when we compare the evolution of $p(t)$ (shown in Figure \ref{fig:p_vs_t_fixedRe}), we note that the cases with $k_p \leq 20$ show a rather erratic behavior, whereas those with greater scale separation are practically indistinguishable and show a smoother behavior. This shows the importance of scale separation to obtain a smooth evolution of $p(t)$ independent of box size effects. 
In principle, the independence of $p_b$ on $k_p$ may be valid only in this range and higher Re, but the situation could be different for lower Reynolds numbers, where a large portion of the energy decay is dissipative.

To address this, we extend the previous study by varying $k_p$ and Re simultaneously. We want to look at the Re dependence of $p_b$ and see if this is independent of the scale separation, even for $k_p \sim 1$, where box size effects are strongly noticeable. For this, we choose three different values of $k_p =5,20 $ and $100$. This choice requires an extremely high resolution in some cases, reaching a box with $N = 4096$.  For each value of $k_p$, we run a small number of simulations varying viscosity. The parameters for these runs are shown in Table \ref{tab:varying_Re_and_kp}. 
\begin{table}
    \centering
    \begin{tabular}{lrrrrr}
    \toprule
    
\toprule
                     Run &  Pm &   $\nu$ &  Re &  $k_p$ &   $N$ \\
\midrule

NH\textsubscript{$kR1$} &      1 &  0.009 &    35 &      5 &  1024 \\
NH\textsubscript{$kR2$} &      1 &  0.006 &    53 &      5 &  1024 \\
NH\textsubscript{$kR3$} &      1 &  0.003 &   107 &      5 &  1024 \\
NH\textsubscript{$kR4$} &      1 &  0.0009 &   358 &      5 &  1024 \\
NH\textsubscript{$kR5$} &      1 &  0.0004 &   805 &      5 &  2048 \\
NH\textsubscript{$kR6$} &      1 &  0.009 &     8 &     20 &  1024 \\
NH\textsubscript{$kR7$} &      1 &  0.006 &    13 &     20 &  1024 \\
NH\textsubscript{$kR8$} &      1 &  0.003 &    26 &     20 &  2048 \\
NH\textsubscript{$kR9$} &      1 &  0.001 &    80 &     20 &  2048 \\
NH\textsubscript{$kR10$} &      1 &  0.0009 &    89 &     20 &  2048 \\
NH\textsubscript{$kR11$} &      1 &  0.0004 &   201 &     20 &  2048 \\
NH\textsubscript{$kR12$} &      1 &  0.009 &     1 &    100 &  4096 \\
NH\textsubscript{$kR13$} &      1 &  0.006 &     2 &    100 &  4096 \\
NH\textsubscript{$kR14$} &      1 &  0.003 &     5 &    100 &  4096 \\
NH\textsubscript{$kR15$} &      1 &  0.001 &    16 &    100 &  4096 \\
NH\textsubscript{$kR16$} &      1 &  0.0009 &    17 &    100 &  4096 \\
NH\textsubscript{$kR17$} &      1 &  0.0004 &    40 &    100 &  4096 \\
\bottomrule
\bottomrule
    \end{tabular}
    \caption{Nonhelical runs for varying viscosity and $k_p$ at fixed Prandtl number Pm $ = 1$.}
    \label{tab:varying_Re_and_kp}
\end{table}

\begin{figure}
    \centering
    \includegraphics[width=\linewidth]{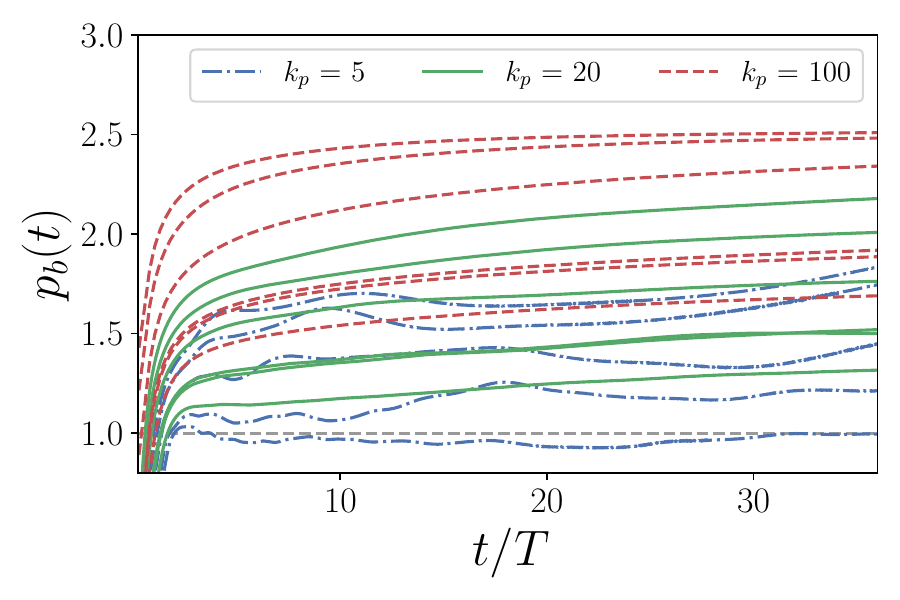}
    \caption{Time evolution of $p_b(t)$ for runs NH\textsubscript{kR}, for different values of Re and $kp = 5$ (blue dash-dotted lines), $kp = 20$ (green solid lines) and $k_p=100$ (red dashed lines).}
    \label{fig:pb_vs_t_kpRe}
\end{figure}

We measure the evolution of $p_b(t)$ for all runs and study its dependence on Re. We take a look at Figure \ref{fig:pb_vs_t_kpRe}, and we note the erratic behavior for the cases with small scale separation. We see that it is not straightforward to determine a time interval to perform a fair measurement between all cases. Despite the erratic behavior, cases with $k_p= 5$ show a plateau between $t/T\approx 15$-$25$. During this time interval, some of the cases with $k_p=20$ and $k_p=100$ show an increasing $p_b(t)$, approaching a possible plateau at later times. For the measurements, we take a narrow interval $t/T = 20$-$25$ to prevent as much bias as possible. 

\begin{figure}
    \centering
    \includegraphics[width=\linewidth]{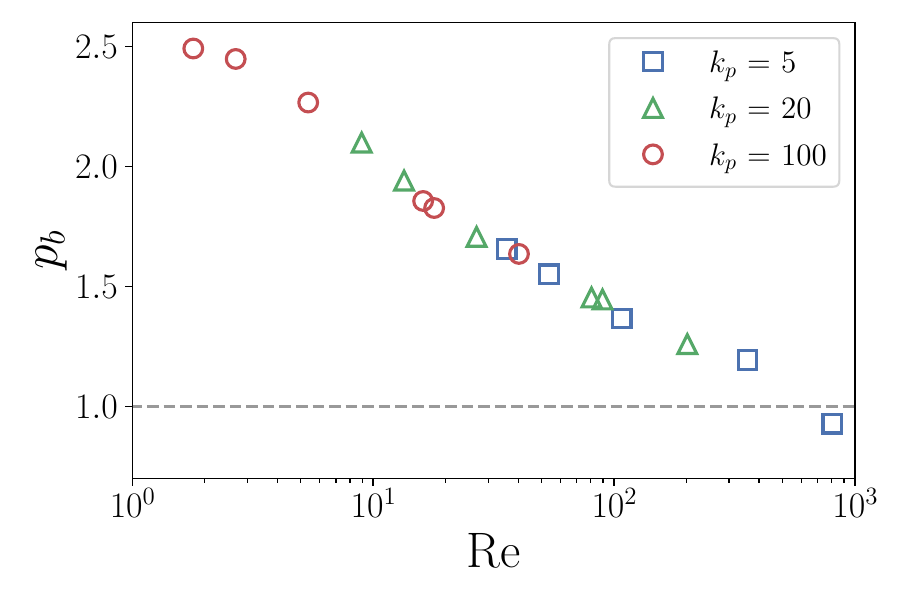}
    \caption{Scaling exponents $p_b$ measured in runs NH\textsubscript{kR}. Blue squares correspond to $k_p=5$, green triangles to $k_p=20$ and red circles to $k_p=100$.}
    \label{fig:p_vs_kpRe}
\end{figure}

In Figure \ref{fig:p_vs_kpRe}, we show the scaling exponents measured for varying Re. We find that $p_b$ follows a clear trend that only depends on Re, without any dependence on the scale separation. Even though an asymptote will be reached at higher values or Re, the range explored is not enough to determine such a value. The increase in computational power is crucial to show this asymptotic trend and to determine the value of $p_b$. Nevertheless, the trend seems to favour the prediction $p_b = 1$, rather than $p_b = 10/7$.

\section{Hyperviscosity and hyperresistivity}
\label{se:hyperviscous}

\begin{figure*}[ht]
\centering
\subfloat[]{
  \includegraphics[width=0.5\linewidth]{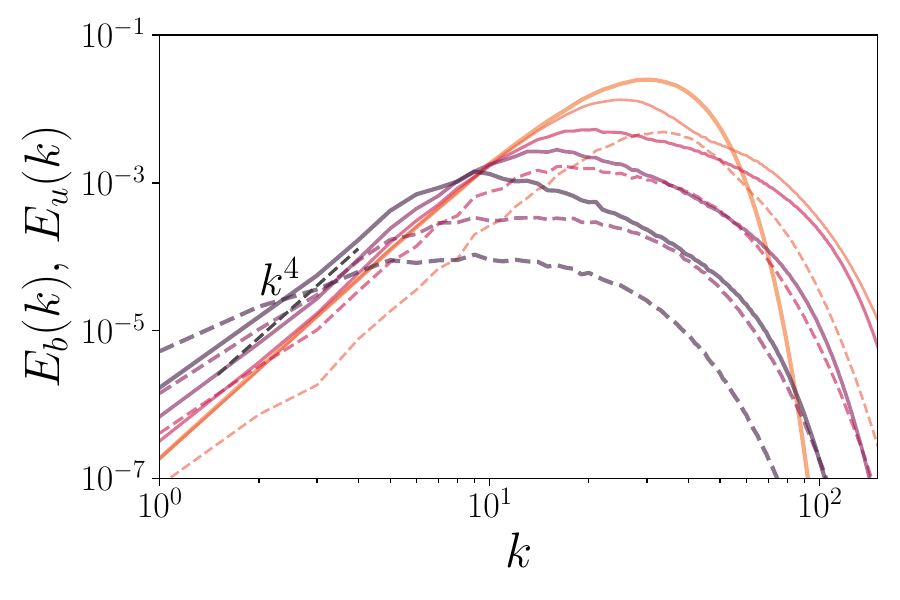}
  \label{fig:hyper_Pm12}
}
\subfloat[]{
  \includegraphics[width=0.5\linewidth]{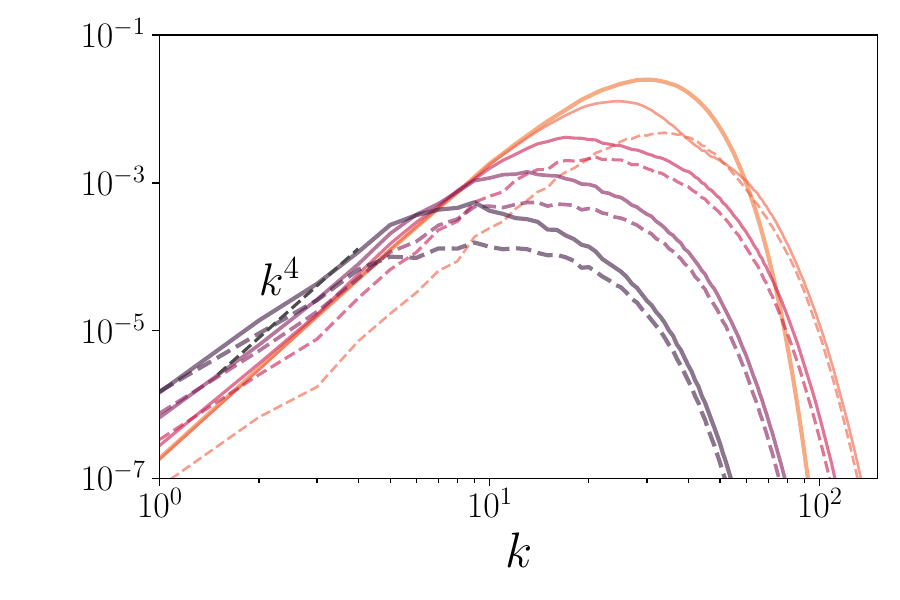}
  \label{fig:hyper_Pm1}
}\\
\subfloat[]{
 \includegraphics[width=0.5\linewidth]{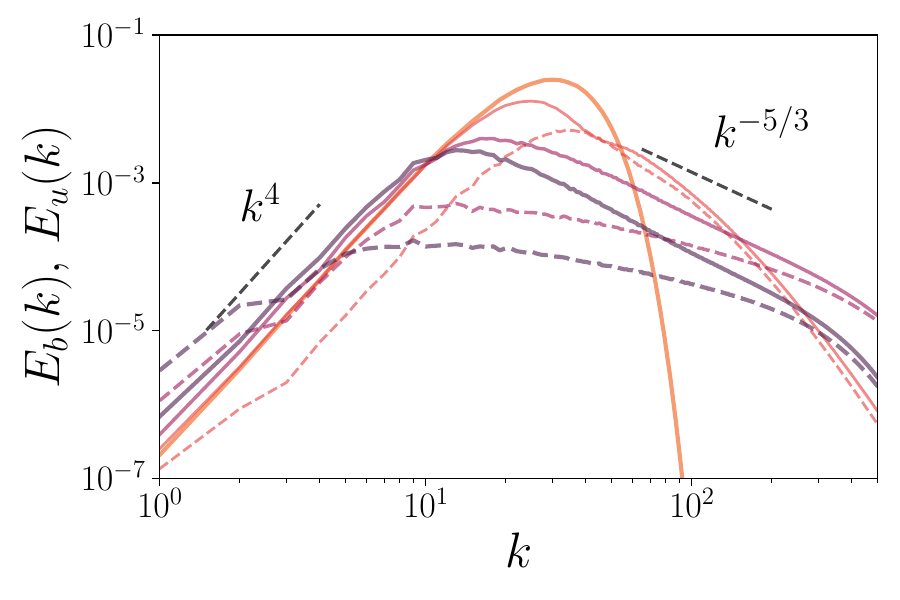}
  \label{fig:hyper_2048}
}
\caption{Spectra evolution of runs (a) NH\textsubscript{hy2}, (b) NH\textsubscript{hy1}, for times $t/T = 0$, $0.63$, $2.5$, $10$ and $40$ and (c) NH\textsubscript{hy3} for times $t/T = 0$, $0.63$, $5$ and $14$. Solid lines indicate magnetic energy spectra and dashed lines indicate kinetic energy spectra. Brighter lines correspond to earlier times.}
\label{fig:hyper_runs}
\end{figure*}

In order to make a fair comparison with previous results in the literature, we run three simulations using hyperviscosity, initialized with the velocity field set to zero. Parameters are reported in Table \ref{tab:hyper_runs}, and the spectra evolution in all three cases are shown in Figure \ref{fig:hyper_runs}.
\begin{table}
    \centering
\begin{tabular}{lrrrr}
\toprule
\toprule
Run &  Pm &    $\nu_2$  &  $k_p$ &   $N$ \\
\midrule
NH\textsubscript{hy1}  &      1 &  $10^{-6}$ &       30 &   1024 \\
NH\textsubscript{hy2} &      12 &  $10^{-6}$ &       30 &   1024 \\
NH\textsubscript{hy3} &      1 &  2$\cdot10^{-10}$ & 30 &   2048 \\
\bottomrule
\bottomrule
\end{tabular}
    \caption{Nonhelical runs using hyperviscosity, initialized with zero velocity field. The Prandtl number is defined as Pm = $\nu_2/\eta_2$.}
    \label{tab:hyper_runs}
\end{table}

In all hyperviscous runs we can observe a slightly stronger inverse transfer than in our standard simulations but still not enough for the peak to move beyond the initial $k^4$ spectrum. In Figure \ref{fig:hyper_2048} we see an inertial range slightly shallower than the $k^{-2}$ found in previous works, closer to the $k^{-5/3}$ scaling. Figures \ref{fig:hyper_Pm1} and \ref{fig:hyper_Pm12} show runs with the same hyperviscosity but with different Pm. The case with higher Pm shows the stronger inverse transfer, supporting the same trend that we observe for standard viscosity.

\section{Comparison with previous literature}
\label{se:comparison_literature}

Some of the features of nonhelical decay that are observed in previous work, namely, the inverse energy transfer or the formation of a weak turbulent spectra $k^{-2}$, are different to the ones found here. Those studies used different codes with slightly different equations from the ones we used in this work. For instance the \texttt{PENCIL} code used in \cite{brandenburg2015nonhelical,park2017inverse,reppin2017nonhelical,bhat2021inverse}, solves compressible MHD, whereas in \cite{zrake2014inverse}, relativistic MHD equations are implemented.

Most of the simulations in recent work are initialized with small or zero kinetic energy. In all cases, the inverse transfer of magnetic energy is stronger than the one we find in this work. In \cite{bhat2021inverse}, an initial kinetic dominated flow is also studied, finding a decay similar to the one found in this work. 

Some minor discrepancies can also be expected in the scaling exponents due to the measurement methods. Additionally, certain numerical aspects such as dealiasing rules, resolution criteria and timestepping procedure can have some impact in the results \cite{beresnyak2019mhd}. Other aspects such as the initial spectra have a relevant influence in the subsequent evolution of the decay \cite{brandenburg2017classes}. Additionally, the use of hyperviscosity and hyperresistivity, which gives a wider inertial range for limited resolution, alters the dissipative mechanisms and the dynamics of the decay \cite{hosking2021reconnection}.

In \cite{brandenburg2015nonhelical,reppin2017nonhelical,bhat2021inverse,park2017inverse}, the \texttt{PENCIL} code is used. The initial magnetic spectra is the same as ours for $k<k_p$, and the kinetic spectra develops a $k^2$ form. In most cases the flow is magnetically dominated at $t=0$ except from one run in \cite{bhat2021inverse} where the magnetic field is subdominant and in \cite{park2017inverse}, where the nonhelical simulation was first driven with a random forcing and relatively small scale separation. In all these works an inverse transfer is observed in the nonhelical case. This transfer is especially strong in \cite{brandenburg2015nonhelical,reppin2017nonhelical}, where the peak of the spectrum goes well past the initial $k^4$ spectrum. This is possibly achieved because of the extremely low values of viscosity, resistivity, hyperviscosity and hyperresistivity. On the other hand, in \cite{bhat2021inverse,park2017inverse}, only standard dissipative terms with moderate values of viscosity are used, obtaining only a moderate inverse transfer. This suggests that the use of hyperresistivity might have a stronger influence on the strength of the inverse transfer than expected. Looking at our results using hyperviscosity in section \ref{se:hyperviscous}, we note that this trend is indeed observed. However, the strength of the inverse transfer is not as great as in these other papers.

In \cite{reppin2017nonhelical}, the results are benchmarked against \textit{Zeus-MP2}, where no inverse transfer is observed. It is argued that this difference is caused by the different numerical integration schemes used (six order finite difference for \texttt{PENCIL} vs. second-order finite difference for \textit{Zeus-MP2}). The authors suggest that the lower order in the numerical integration of the latter, adds a numerical dissipation that might affect the evolution of the magnetic field. A previous work studied the differences between Snoopy, \texttt{PENCIL} and \textit{Zeus-MP2} \cite{fromang2007mhd}. In this work, the authors conclude that the transport properties are not affected severely by numerical dissipation. No parameters are given for the \textit{Zeus-MP2}, hence the comparison that we can do is limited. Nevertheless, in our simulations, we find an opposite trend at increasing Pm, and a much slower growth of the magnetic coherence length $L_b(t)$ than that found in \cite{reppin2017nonhelical}. Even though the numerical aspects might explain some of the differences observed, we believe that these differences are substantial and need further exploration. 

In any case, the discussions given in \cite{hosking2021reconnection} and the arguments and data presented in \cite{bhat2021inverse}, suggest that the inverse transfer of magnetically dominated flows should be stronger than those initially in equipartition. That is also different to what we observe numerically.

In \cite{zrake2014inverse}, the relativistic MHD code \texttt{MARA} was used (see \cite{zrake2011numerical} for details, especially for the Godunov finite-volume integration scheme used and the inherent numerical dissipation). This code has some differences with the rest of the codes mentioned in this section, but it also shows a nonhelical inverse transfer with a noticeable growth of the integral scale over time. It is not mentioned how the kinetic field is initialized in this simulation.

Finally, in \cite{hosking2021reconnection} the Snoopy code is used \cite{lesur2015snoopy}. This code is the most similar to the one we use, since it is a pseudospectral code that solves incompressible MHD in a box of size $2\pi$ with a 2/3 dealiasing rule. The only difference with our code is that Snoopy uses a third-order Runge-Kutta scheme for the timestepping procedure. In this work, the authors implement viscosity and hyperviscosity. The kinetic flow is initialized to zero and a clear inverse transfer is observed in the nonhelical hyperviscous case. The scaling exponents measured using the Snoopy code are in reasonable agreement with ours. This is interesting to note, since the authors use a measurement method where some biases of the log-log fit are overcome. 

We performed a last comparison with the Snoopy code, given that our code uses only a second-order timestepping procedure. We repeated runs NH\textsubscript{p2} and NH\textsubscript{hy1} with a timestep ten times smaller than in the original runs, and we checked that results were stable and that the choice of timestep did not introduce any undesired effect. A possible source for disagreement could be that other invariants such as cross-helicity are set different initially. It has been shown in previous work that cross-helicity can quench triadic interactions producing forward transfer, creating asymmetries that favour inverse transfer \cite{linkmann2016helical}. Even though net cross-helicity vanishes, differences in the local structures can have an impact in the inverse transfer and give place to a different phenomenology. More evidence studying the influence of cross-helicity may clarify if this is relevant to explain the discrepancies.

We can conclude that after an extensive check, the decaying rates we measured in this work are in reasonable agreement with previous literature. This suggests that the discrepancies we find in terms of the inverse transfer are not for any obvious
reasons such as coding errors. Since the underlying mechanisms for the inverse transfer in nonhelical MHD are not yet clear, we believe that further analysis is needed. Either numerical or physical aspects that might seem subtle, could strongly affect the evolution of the magnetic field and in particular, the inverse energy transfer in the nonhelical case.

\section{Discussion and conclusions}

In this work we have explored the decay of helical and nonhelical MHD turbulence using fully resolved DNS in a wide range of parameters. We find a present but weak nonhelical inverse transfer of magnetic energy, compared to the one found in recent literature \cite{brandenburg2015nonhelical,reppin2017nonhelical,hosking2021reconnection,zrake2014inverse}. Nevertheless, we found that increasing Prandtl number enhances this inverse transfer, especially in the kinetic field. This is opposite to the trend found in \cite{reppin2017nonhelical}, where increasing Pm turns the inverse transfer less efficient. This difference is possibly related to the subtleties involved in different numerical implementations of the MHD equations that might affect strongly the mechanisms of nonhelical inverse transfer.

We also measured the helical and nonhelical decay rate $E_b \sim t^{-p_b}$ for different parameters. We note that a careful numerical approach is necessary for measuring these values. Especially, due to the closeness of the different theoretical predictions ranging from $p_b \approx 0.5$-$0.7$ in the helical case and $p_b \approx 1$-$1.5$ in the nonhelical case. We report the trends observed for $p_b$ at varying Prandtl number, varying Reynolds number and varying scale reparation $k_p$. We find that $p_b$ decreases for increasing Pm and increasing Re, producing a shallower magnetic decay in both helical and nonhelical cases. Furthermore we find that in the helical case, the decay follows a functional form $p_b \approx 0.6 + 14/$Re.  

We find that the behavior of the large scales is affected by scale separation in the nonhelical case (the helical case is not studied). A small scale separation shows an erratic evolution of the scaling exponent and the evolution of the subinertial spectra $E_b \sim k^4$. Nevertheless, our numerical results show that the measured values of $p_b$ are not strongly dependent on $k_p$.

In \cite{bhat2021inverse} and \cite{hosking2021reconnection}, the authors suggest that flows in equipartition $U\sim B$ show a weaker inverse transfer and a steeper magnetic decay. However, we do not find strong differences between these two cases, either in the steepness of the decay or the amount of inverse transfer. Still, the magnetically dominated case shows a slightly stronger inverse transfer than the case in equipartition, which is in line with the above mentioned predictions, but not as strong as in other works in literature.

Finally we comment on the similarities and differences between the observed nonhelical inverse transfer in our results and those in recent literature. A strong inverse transfer has been observed using three different codes that use different equations, different numerical techniques, and different fields initialization. We believe that further work is needed to understand the reason for such differences. We made sure that our simulations satisfy adequate spatial and temporal resolution requirements. This is something these other studies
were more lenient in regards. Other properties such as the compressibility present in the \texttt{PENCIL} code were also suggested in \cite{reppin2017nonhelical} as a possible mechanism to enhance the inverse transfer due to the form of the kinetic spectra at low wavenumbers. Nevertheless, the strong inverse transfer observed in \cite{hosking2021reconnection} using incompressible turbulence indicate that the source of the discrepancy might not be related to this. We argue that the initial cross-helicity and its subsequent evolution can be the source of the observed discrepancies. Last, we run a small number of hyperviscous and hyperresistive runs, to verify if the inverse transfer emerges when the inertial range is sufficiently wide. The comparisons shows a slightly stronger inverse transfer but not enough to see the peak of the spectrum moving past the initial $k^4$ slope.

\section*{Acknowledgements}
We would like to thank Moritz Linkmann for useful discussions that were helpful to establish some criteria for this investigation. This work used the ARCHER2 UK National Supercomputing Service (https://www.archer2.ac.uk). J.C.F. was supported by the Secretary of Higher Education, Science, Technology and Innovation of Ecuador (SENESCYT) and A.A. was supported by the University of Edinburgh. A.B. acknowledges partial funding from the U.K. Science and Technology Facilities Council.


\bibliography{PRE_MHD_decay.bib}

\end{document}